\documentclass[sigconf]{acmart}

\usepackage{array}
\usepackage[utf8]{inputenc}
\usepackage{comment}
\usepackage{color}
\usepackage{graphicx}
\usepackage{caption}
\usepackage{subcaption}
\usepackage{amsmath}
\usepackage{multirow}
\usepackage{algorithm,algorithmicx}
\usepackage{flushend}
\usepackage[noend]{algpseudocode}

\algrenewcommand\algorithmicforall{\textbf{foreach}}
\algrenewcommand\algorithmicindent{.8em}

\makeatletter
\def\BState{\State\hskip-\ALG@thistlm}
\makeatother
\renewcommand\footnotetextcopyrightpermission[1]{} 
\newenvironment{packed_itemize}{
\begin{list}{\labelitemi}{\leftmargin=2em}
\vspace{-1.5pt}
 \setlength{\itemsep}{2pt}
 \setlength{\parskip}{2pt}
 \setlength{\parsep}{2pt}
}{\end{list}}
\title{Don't Mine, Wait in Line: Fair and Efficient \\ Blockchain Consensus with Robust Round Robin}

\author{Mansoor Ahmed-Rengers}
\affiliation{%
  \institution{University of Cambridge}
  \city{Cambridge}
  \country{UK}}
\email{mansoor.ahmed@cl.cam.ac.uk}

\author{Kari Kostiainen}
\affiliation{%
  \institution{ETH Zurich}
  \city{Zurich}
  \country{Switzerland}}
\email{kari.kostiainen@inf.ethz.ch}

\begin{document}
\maketitle


\section*{Abstract}

Proof-of-Stake systems randomly choose, on each round, one of the participants as a consensus leader that extends the chain with the next block such that the selection probability is proportional to the owned stake. However, distributed random number generation is notoriously difficult. Systems that derive randomness from the previous blocks are completely insecure; solutions that provide secure random selection are inefficient due to their high communication complexity; and approaches that balance security and performance exhibit selection bias. When block creation is rewarded with new stake, even a minor bias can have a severe cumulative effect.

In this paper, we propose \emph{Robust Round Robin}, a new consensus scheme that addresses this selection problem. We create reliable long-term identities by bootstrapping from an existing infrastructure, such as Intel's SGX processors, or by mining them starting from an initial fair distribution. For leader selection we use a deterministic approach. On each round, we select a set of the previously created identities as consensus leader candidates in round robin manner. Because simple round-robin alone is vulnerable to attacks and offers poor liveness, we complement such deterministic selection policy with a lightweight endorsement mechanism that is an interactive protocol between the leader candidates and a small subset of other system participants. Our solution has low good efficiency as it requires no expensive distributed randomness generation and it provides block creation fairness which is crucial in deployments that reward it with new stake.


\section{Introduction}
\label{sec:intro}

\begin{figure}[t]
  \centering
  \includegraphics[width=0.85\linewidth]{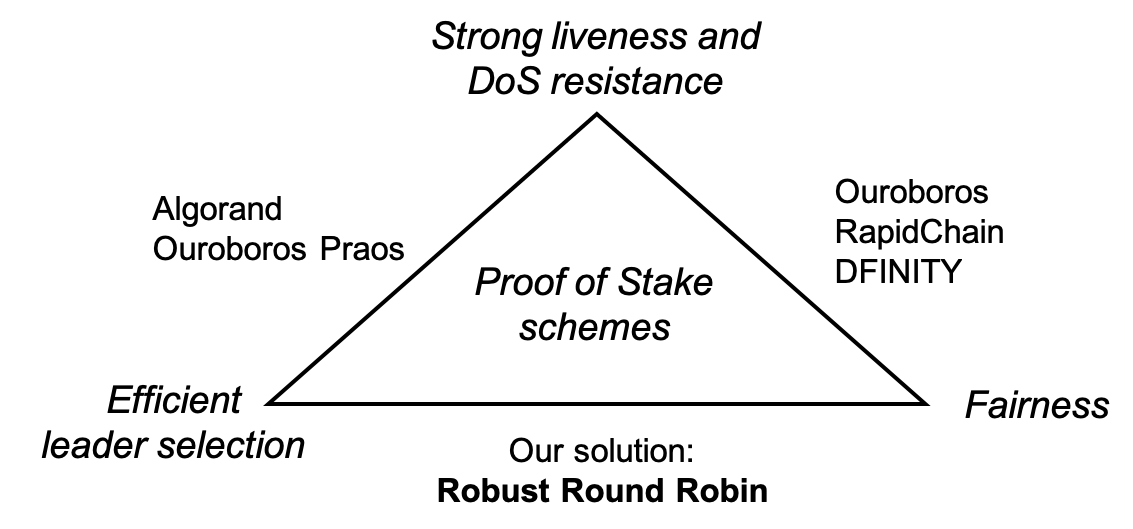}
  \caption{The state-of-the-art PoS systems choose consensus leaders randomly which enables good liveness and DoS resistance. Some such schemes use distributed randomness generation that is efficient but can be biased and therefore such solutions do not provide fairness. Other schemes use unbiased distributed randomness generation with high communication cost. Our solution, \emph{Robust Round Robin}, is fair and efficient, but provides weaker DoS resistance.}
  \label{fig:tradeoff}
  \vspace{-5pt}
\end{figure}

Any decentralized digital currency requires a consensus mechanism to prevent double spending. Bitcoin~\cite{bitcoin} leverages Proof of Work (PoW)~\cite{dwork-1992} and economic incentives to achieve consensus in a permissionless setting, assuming that the adversary does not control majority of the computing power and messages are delivered sufficiently reliably~\cite{garay-2015}. Experience from the last ten years has shown that it is indeed possible to realize a digital currency without a trusted authority. However, Bitcoin also has serious limitations: its throughput is low (7 tps), its latency is high (60 minutes) and most importantly it wastes huge amounts of energy. The estimated energy consumption of all Bitcoin miners is comparable to a medium size country~\cite{digiconomist}.

Recently, several alternative permissionless blockchain consensus schemes have been proposed. Proof of Stake (PoS) is arguably the most prominent approach that avoids the above energy waste. The basic idea in most PoS systems is to \emph{randomly choose}, on each round, one of the system participants as a consensus leader that extends the chain with a new block. Selection is performed such that the probability of being chosen as the leader is proportional to the owned stake like coins.

However, secure random selection in a distributed setting is difficult. The initial PoS schemes that derive randomness from the previous blocks are susceptible to grinding attacks and completely insecure. Recent PoS schemes like Ouroboros~\cite{Ouroboros2017} or DFINITY~\cite{dfinity} that use secure random beacon protocols have high communication and computation complexity and are therefore inefficient. PoS systems like Algorand~\cite{algorand} and Ouroboros Praos~\cite{praos} enable more efficient random selection, but such schemes are susceptible to selection bias.
When block creation is rewarded with new stake, as is a common practice, even a small bias can have a large cumulative effect. For example, in Algorand-style leader selection an adversary that controls $\alpha=0.33$ of stake when the system has 1,000 units of stake, collects 36\% of block rewards and new stake. Once the system has 10,000 units of stake, the adversary controls majority of the stake and collects more than 70\% of the rewards.

Usage of trusted execution environments (TEEs), such as Intel's SGX, have also been proposed as a solution to the leader selection problem. Proof of Elapsed Time (PoET)~\cite{sawtooth} is an example system where attested enclaves wait a random time and the enclave that finishes first becomes the leader. The drawback is that a compromised SGX processor wins the leader selection arbitrarily often, and thus participants have an incentive to attack their own platforms. Several successful attacks against SGX have been recently demonstrated~\cite{van2018foreshadow,intel-vulnerability,sgx_sidechannel,sgx_kari_sidechannel}.

\paragraph{Our solution}
In this paper, to addresses the leader selection problem, we propose a new permissionless blockchain consensus scheme called \emph{Robust Round Robin}. We establish reliable and long-term identities and record the enrollment of each identity to the ledger. The number of identities controlled by each participant in the system is limited to their stake and we propose two concrete ways to establish identities, and thus two notions of stake.

Our first identity creation mechanism is bootstrapping from existing infrastructures. As an example infrastructure, we use Intel's SGX processors and attestation service (IAS) such that the stake of each participant is the number of SGX processors he controls. Although we instantiate our solution using SGX, we emphasize that our approach is not limited to SGX, but similar identities could be bootstrapped also from other infrastructures such as mobile subscriptions or credit cards. Our second identity creation mechanism is ``mining'' the identities starting from an initial fair distribution. In this approach, the identities themselves function as stake. The first approach applies to partially-decentralized setting where the consensus is permissionless but the infrastructure providers, like Intel, needs to be trusted (for attestation). The second approach applies to fully-decentralized setting similar to Bitcoin.

Our solution performs \emph{deterministic} leader candidate selection. We assign an age to each identity and place them into a queue in the order of decreasing age. Our notion of age is the number of rounds since the enrollment of the identity or its previous successful block creation. Once a chosen leader candidate creates a block successfully, its age becomes zero and it moves to the end of the queue again, essentially achieving \emph{round-robin} candidate selection. 

Because such simple round-robin selection is vulnerable to attacks and provides poor liveness, we complement it with a lightweight endorsement mechanism. On each round, we sample a small subset of other identities as endorsers. Each deterministically-chosen leader candidate runs a simple protocol with the endorsers and the candidate that receives the required quorum of confirmations becomes the leader to create a new block.  In rare cases, more than one candidate may be chosen, or more than one block created by the same leader, but the probability of such events on multiple successive rounds reduces exponentially and therefore forks remain shallow. The adversary may bias endorser selection, but that does not enable attacks like double spending or increase his rewards. We call our solution \emph{Robust Round Robin}, in contrast to simple and insecure standard round robin.

The main benefits of our solution compared to other PoS systems are fairness and efficiency. As highlighted in Figure~\ref{fig:tradeoff}, solutions like Algorand~\cite{algorand} and Ouroboros Praos~\cite{praos} suffer from selection bias which can have large cumulative effect. In our solution leader selection is based on deterministic schedule and thus fair. Solutions like Ouroborous~\cite{Ouroboros2017} and DFINITY~\cite{dfinity} require expensive protocols to establish unbiased randomness periodically. Our lightweight endorsement protocol is simple and efficient. In contrast to previous TEE solutions like PoET~\cite{sawtooth}, participants gain no advantage by compromising their own platforms and in this regard our solution is resilient to TEE compromise.

Deterministic leader selection has also drawbacks in contrast to randomized selection, as shown in Figure~\ref{fig:tradeoff}. Because the selection schedule is predictable, our solution can be more susceptible to denial-of-service (DoS) attacks that target the next leader. Another concern is an adversary that owns several oldest identities and therefore controls block creation on several successive rounds. Such adversary could prevent transaction processing from targeted users temporarily. Although such DoS attacks cannot be prevented fully, we outline ways to make them difficult to deploy in practice.

The performance and scalability of our solution is comparable to recent PoS schemes. Users can consider transactions safely confirmed once they are extended by a small number of blocks (e.g., $d=6$ or $12$). Since our endorsement protocol is simple, rounds can be set short (e.g., 5 seconds in our experiments) which gives one or half a minute transaction latency and throughput of 1500 tps. The per-round communication and computation complexity is constant and small (e.g., approximately 100 messages per round).

\paragraph{Contributions}
To summarize, our paper makes the following contributions:
 
\begin{packed_itemize}
    \item \emph{Selection bias analysis.} Our analysis identifies that even a small bias in leader selection can have drastic consequences in systems where block creation is rewarded with new stake. To the best of our knowledge, this paper is the first to explain this problem in detail.

 	\item \emph{New consensus scheme.} We propose \emph{Robust Round Robin}, a novel consensus scheme, where deterministic leader candidate selection is complemented with a lightweight interactive endorsement protocol. The main benefits of our approach are efficiency and fairness.

 	\item \emph{Analysis and experiments.} We analyze the security of our solution and estimate its performance using experiments in a custom-built peer-to-peer network.
\end{packed_itemize}

\paragraph{Outline.}
The rest of the paper is organized as follows. Section~\ref{sec:background} provides background on blockchain consensus and motivates our work. Section~\ref{sec:overview} presents an overview of our solution. Section~\ref{sec:identity} details identity creation and Section~\ref{sec:operation} system operation. Section~\ref{sec:analysis} provides security analysis and Section~\ref{app:evaluation} performance evaluation. Section~\ref{sec:discussion} presents a discussion, Section~\ref{sec:related-work} reviews related work and Section~\ref{sec:conclusion} concludes the paper.


\section{Background and Motivation}
\label{sec:background}




The collective puzzle solving in PoW provides randomized leader selection, where the selection probability is proportional to the amount of performed work. Most PoS systems mimic such random selection and choose a leader with a probability that is proportional to the owned stake. In this section we review previous schemes and their limitations.

\paragraph{Naive random selection.} 
The first PoS proposals~\cite{peercoin,pos_history} suggested a simple technique where the hash of the previous block functions as a ``random'' seed for leader selection on the next round. However, this approach is vulnerable to \emph{grinding attacks}, where the leader of the previous rounds tries different block candidates (e.g., by sampling from the pool of pending transactions) and picks the block that gives him an advantage in leader selection on the next round. By iterating through many candidate blocks he can pick one that makes him the leader on the next round as well.

Another simple approach is to run a bias-resistant \emph{random beacon} protocol among all participants with stake. Random beacon is a distributed protocol that generates a new random value periodically. The main drawback of this approach is that such protocols traditionally have high communication and computation cost. For example, the complexity of Cachin's random beacon~\cite{cachinrandomgen} is $O(n^3)$ which means that performing repeated random selection among all participants is either very expensive or completely infeasible.

\paragraph{Sophisticated random selection.}
Recent research has suggested more efficient random beacons, both as standalone protocols and as part of PoS blockchain systems.

RandHerd~\cite{syta2017randhound} is a standalone random beacon that leverages publicly verifiable secret sharing (PVSS) and collective signing (CoSi) to produce unbiased and unpredictable random values among large set of participants. RandHerd divides all participants into smaller committees of size $c$. A required threshold of participants from each committee contributes to the output random value. The per round complexity of RandHerd is reduced to $O(c^2 log(n))$. The main problem for permissionless blockchains is an expensive initialization routine where participants are divided into groups which takes several minutes. Such slow reconfiguration should be repeated when new participants join or leave the system.

Ouroboros~\cite{Ouroboros2017} is a PoS system with a built-in random beacon. Ouroboros randomly samples a committee that runs PVSS-based protocol with complexity of $O(n^3)$. Since such protocol is executed only once per epoch, the high cost is amortized over several rounds. However, the main drawback of this solution is that it requires committees with honest majority which means that committees of \emph{thousands of participants} must be used to ensure their honest majority across the entire system lifetime which makes the protocol very expensive. Moreover, Ouroboros requires synchronous communication which is difficult to achieve in large peer-to-peer networks.

In Algorand~\cite{algorand}, random values are derived using verifiable random functions (VRFs)~\cite{vrf-micali-99}. On each round, the chosen leader computes the next random value using a VRF and the previous random value. A publicly verifiable proof $\pi$ of this computation is added to the block. VRF-based selection is efficient, but the main problem is that such approach is not bias-resistant. The chosen leader may bias the protocol output, e.g., by skipping his turn.

\begin{figure}[t]
  \centering
  \includegraphics[width=0.9\linewidth]{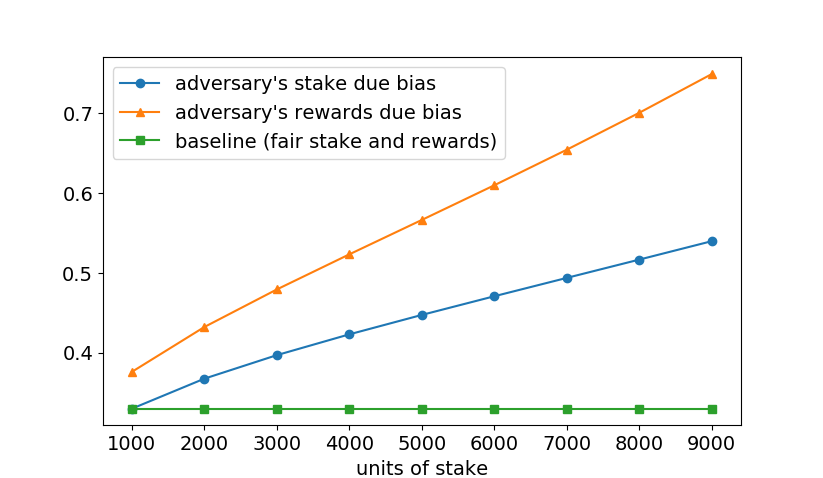}
  \caption{Example of the cumulative effect of minor selection bias
  in a system where block creation is rewarded with new stake. Triangles show the adversary's block creation rate and rewards increase. Circles show the adversary's stake increase. Squares represent the baseline ($\alpha=0.33$) of fair rewards and stake without bias.
  }
  \label{fig:cumulative}
  \vspace{-10pt}
\end{figure}

To illustrate the effect of such selection bias, consider an example system where multiple leader candidates with priorities are chosen. This approach is used in most PoS systems, because choosing only one leader prevents the system from proceeding in case it is offline or otherwise unable to communicate. Assume an adversary that controls a fraction $\alpha=0.33$ of stake. On the average, every 9th round the leader candidate with the first and second priority both belong to the adversary, every 27th rounds this is the case for the top three priority candidates, and so on. The adversary can now choose which one of these leader candidates to use and pick the one that gives the most advantageous value for next selection. While such bias can be relatively small, its effect will cumulate when system participation is incentivized by providing rewards like new stake to the chosen leader (block creator), as is a common practice in blockchain systems. Figure~\ref{fig:cumulative} shows an example starting from 1,000 units of stake. Due to the above bias, the adversary creates blocks at slightly higher rate ($\approx$ 0.36) and thus his share of stake increases. By the time the system has 10,000 units of stake, the adversary controls majority of the stake and creates over 70\% of the blocks, at which point the system can no longer be secure. 

Ouroboros Praos~\cite{praos} is another PoS scheme that also leverages VRFs for leader selection. Similar to Algorand, selection can be biased by the adversary. DFINITY~\cite{dfinity} and RapidChain~\cite{rapidchain} are further examples of recent PoS schemes that performs unbiased leader selection with significant communication cost. We review such solutions, and their limitations, in Section~\ref{sec:related-work}.

\paragraph{Selection using TEEs.} In PoET~\cite{sawtooth}, the consensus participants are attested SGX enclaves and the enclave that finishes randomized waiting first is chosen as the leader. Assuming that SGX ensures code integrity, this approach enables secure leader selection.
The main drawback is that participants have an incentive to break one of their own SGX processors which allows the participant to win the leader selection arbitrary often. SGX was designed to protect enclaves against malicious software and but not against physical attacks. Additionally, recent research has demonstrated that software-only attacks like Foreshadow~\cite{van2018foreshadow,intel-vulnerability} that leverage the Meltdown vulnerability~\cite{Lipp2018meltdown} can extract attestation keys from SGX processors, essentially enabling a full break of SGX. Developing schemes that detect processors that win statistically ``too often'' may be possible, but eliminating all bias is difficult.

\paragraph{Requirements.}
Given these limitations of previous solutions, our goal is to design a permissionless blockchain consensus scheme that meets the following requirements. 

\begin{packed_itemize}
	\item \emph{Fairness.} Our solution should ensure leader selection fairness. As explained above, even a relatively small bias in leader selection can have severe cumulative effects when combined with block creation rewards. 
    
    \item \emph{Efficiency.} Our solution should be efficient. In particular, we want to avoid complicated leader selection protocols that have high communication cost. 
    
    \item \emph{Tolerance to TEE compromise.} If TEEs are used, the adversary should not gain advantage by compromising protections on his \emph{own} processors. Such property enables robust deployments in practice.
\end{packed_itemize}



\section{Robust Round Robin Overview}
\label{sec:overview}

In this section, we provide an overview of our solution that we call \emph{Robust Round Robin}. We start by listing our assumptions. After that we explain our main ideas, discuss challenges, and finally provide an overview of the solution.

\subsection{Assumptions}

We assume that the blockchain is used for a digital currency or another application that enables rewards that incentivize behavior similar to monetary rewards. The blockchain is used to record and order transactions, but our solution is agnostic to their type. 

We consider two \emph{trust models}. The first is a \emph{partially-decentralized} setting, where the blockchain consensus is maintained by a permissionless set of participants, but we rely on the integrity of an existing infrastructure (e.g., that Intel manufactures processors and runs the SGX attestation service correctly). Our second trust model is a \emph{fully-decentralized} setting with no trusted entities similar to Bitcoin and most other permissionless blockchains.

We consider an adversary that controls a significant fraction $\alpha$ of stake (e.g., $\alpha = 0.33$). Our adversary model is \emph{non-adaptive} in the sense that the adversary cannot arbitrarily choose for every time window which computing platforms or users' key pairs he controls.\footnote{We note that some previous works like Algorand and Ouroboros consider a stronger \emph{fully-adaptive} adversary~\cite{algorand,Ouroboros2017} that can freely choose controlled participants for each time window. Our take is that such a fully-adaptive adversary is interesting, and worth studying, but often not realistic. In practice, platform compromise is hard to detect and repair. Furthermore, a compromise of one computing platform does not mean that another is no longer in control of the adversary. For these reasons, we focus on non-adaptive adversaries in this work.}
If TEEs are used, we assume that the adversary can extract secret keys, such as attestation keys, and modify attested enclave code on all of his \emph{own} processors.

We assume that the participants communicate over a peer-to-peer network. Within each time window $t$, each participants is able to communication with all other participants except a small fraction $\beta$ (e.g., $\beta = 0.05$). This model is motivated by previous studies on the Bitcoin network where most, but not all, nodes receive broadcasted messages within a delay that can be easily estimated~\cite{bitcoin-stats,Decker-P2P13}. 
Finally, we assume that participants have loosely synchronized clocks.

\subsection{Identity Creation}
\label{sec:overview:idea}

We create of long-term and reliable \emph{identities} and record the enrollment of each identity into the blockchain. 
This approach can be used with different types of stake and we describe two concrete ways to establish identities, and thus two notions of stake.

The first is bootstrapping identities from an \emph{existing infrastructure}. We use Intel SGX processors as an example infrastructure to instantiate our solution because its attestation service provides the needed interface to implement our solution. However, we emphasize that our solution is not limited to SGX and similarly identities could be bootstrapped from other infrastructures like mobile subscriptions, credit cards, passports or other TEEs~\cite{gradient}. We discuss this further in Section~\ref{sec:discussion}. When SGX is used, the stake of each participant is the number of enrolled SGX processors he controls. This approach works in the partially-decentralized setting where trust on Intel, or similar infrastructure provider, is required.

Our second way to create identities is to ``mine'' them starting from an initial fair distribution. That is, successful block creation is rewarded with new identities. In this approach, the controlled identities themselves function as the stake. This approach works in the fully-decentralized setting. The initial fair distribution of identities can be created using PoW. 

\subsection{Starting Point: Deterministic Selection}
\label{sec:overview:idea}

The starting point of our solution is \emph{deterministic selection}. We assign an \emph{age} to each identity such that the age refers to the number of rounds since its recorded enrollment or previous block creation event, and we place all enrolled identities to a virtual \emph{queue} that is sorted in the decreasing order of age.\footnote{Although in this paper we use the age of each identity, we note similar round-robin selection could be realized also through other means such as selecting identities in alphabetical order.} Once an identity creates a block successfully, its age becomes zero again and it moves back to the end of the queue (i.e., round-robin selection). Because such selection schedule is deterministic, the adversary cannot bias it.

\paragraph{Security and liveness challenges}

Given such identities and deterministic selection approach, perhaps the simplest solution would be to select only the single oldest identity as the eligible leader on each round and define that a valid chain cannot skip rounds. 
While such solution would have little ambiguity about the correct leader on each round, the solution would be completely impractical, since
the entire system would stop proceeding when the single selected leader is offline or otherwise unable to communicate sufficiently fast (i.e., poor \emph{liveness}). Thus, we focus on solutions where multiple oldest leader candidates, with priorities in the order of their age, are selected and a valid chain is allowed to skip rounds. Such a system is able to produce a new block on each round with high probability and proceed even from the rare scenarios where all selected candidates are unavailable to communicate, like a temporary large-scale network outage.


Since we allow multiple leader candidates for each round, we must define which chain branch is considered valid in case more than one eligible leader candidate creates a block (i.e., the chain forks). Another simple solution would be to parse the chain starting from the beginning and on each fork favor the oldest leader candidate. However, this approach would allow so called \emph{re-writing the history attacks} where the adversary intentionally skips block creation on the round where he is the oldest leader candidate, but after a long time publishes a blocks that creates a fork deep in the chain. 

To avoid such attacks, we adopt the common ``longest chain'' policy where the branch with the most valid blocks is valid. Given this definition, we have two remaining design challenges to consider. The first challenge, \emph{deep forks}, is about security. Since enrollment of new identities is open (permissionless), the adversary could enroll, say, 50 identities successively. Once these identities become the oldest, they would be chosen as the leader on 50 successive rounds. On each round, the adversary could extend one chain branch with a new block that he broadcasts to the network immediately and another block on a separate branch which will be published later causing a fork that is 50 blocks deep. Both branches would be equally long and thus valid. This attack is an instance of the \emph{nothing-at-stake} problem that is a common challenge in PoS systems.

The second challenge, \emph{inactive identities}, is about liveness. We propose a scheme where participants establish long-lived identities, but it would be unrealistic to assume that all participants stay active in the system forever. 
Old and inactive identities could cause extended periods where the system is unable to produce blocks. 

\subsection{Final Solution: Robust Round Robin}

To address the above two challenges (i.e., deep forks and inactive identities), we complement the simple and deterministic round-robin selection with a \emph{lightweight leader endorsement} mechanism. 

On each round, a small set of oldest identities are chosen as leader candidates. Additionally, we \emph{randomly} sample a subset of recently active identities to serve as endorsers. The sampling is based on a random seed that is updated for each new block using verifiable random functions (VRFs) similar to~\cite{algorand,praos}. Each leader candidate performs a simple interactive protocol with the endorsers. In this protocol, a candidate proposes a block and the endorsers confirm the block from the oldest candidate they observe. The leader candidate that receives the required quorum of $q$ confirmations from the endorsers, is chosen as the leader to extend the chain with a new block. 

Essentially, the endorsers act as \emph{witnesses} and vouch that (1) the candidate they confirmed was the oldest active on that round and (2) the candidate committed to extend the chain with a specific block. Such endorsement guarantees that, with high probability, only one block from one leader is produced on each round and that adversaries cannot go back in time to re-write the history. We call our solution \emph{Robust Round Robin} in contrast to simple and insecure standard round robin.

Using random sampling for endorser sampling may sound contradictory to our previous reasoning. However, our key observation is that although sampling can be biased by the adversary, in our solution it brings no advantage, such as increased rewards or possibility of double spending
(see Section~\ref{sec:analysis} for security analysis).

Besides preventing deep forks, the secondary purpose of the endorsement mechanism is to track active identities. The leader that receives the required quorum of confirmations includes the received confirmations to the created block. By parsing the chain, it becomes possible to verify which identities have activity in the form of confirmation messages, and inactive identities can be excluded from leader candidate and endorser selection.

In the next two sections we describe our solution in more detail. Section~\ref{sec:identity} explains identity creation and Section~\ref{sec:operation} details system operation.


\section{Identity Creation}
\label{sec:identity}

In this section we describe two ways to establish identities for our solution: bootstrapping from existing infrastructures and mining starting from an initial fair distribution.

\subsection{Bootstrap from Existing Infrastructures}
\label{sec:scifer:sgx}

Trusted Execution Environments (TEEs) like Intel's SGX~\cite{sgxwebpage} enable execution of \emph{enclaves} in isolation from any untrusted software. For our solution, the most relevant part of SGX is its attestation protocol where a remote entity can verify that specific enclave code runs on a genuine SGX processor. The attested processor signs a statement over the enclave measurement, which was recorded during its initialization. The verifier forwards the signed statement to Intel Attestation Service (IAS), an online service run by Intel, that sends back a signed attestation evidence. 
SGX attestation uses group signatures and it is anonymous, in the sense that it does not identify the attested hardware platform~\cite{EPIDstandard}. However, SGX attestation supports \emph{linkable} mode that allows the remote verifier to test if the currently attested processor has been previously attested without identifying it. 

We leverage the linkable attestation mode for bootstrapping identities. As identities we use public keys of key pairs that are generated inside enclaves. We bind these keys to the attestation protocol and save the attestation evidence, signed by IAS, to the blockchain. Given such evidence, anyone can verify that the same processor is enrolled \emph{at most once}. 

Importantly, our solution \emph{does not} require enclave data confidentiality or execution integrity. (We use sealing to protect the IAS access credential, but its secrecy is not relevant for consensus.) Thus, our system tolerates adversaries that can compromise their own processors. 

\paragraph{Initialization.}
A new blockchain can be initialized by any, potentially untrusted, entity that we call chain creator. The creator registers with Intel and obtains an access credential $c_a$ for IAS. At registration, the creator specifies that linkable mode of attestation is used (see Appendix~\ref{app:sgx} for details).

The creator chooses $n_0$ platforms as the initial system members. These platforms install enclave code that creates an asymmetric key pair, seals the private key $sk_i$, and exports the public key $pk_i$. The creator performs a remote attestation on each of the selected platforms. During attestation, each platform supplies a hash of $pk_i$ as the \textsf{USERDATA} to be included as part of the \textsf{QUOTE} structure $Q_i$. If the attestation is successful, IAS signs $Q_i$ that includes a pseudonym $p_i$ for the attested platform. The attested enclaves send their public keys $pk_i$ to the creator. 

The creator checks that the public keys match the respective hashes reported in each \textsf{QUOTE} structure $Q_i$ and that all attested platforms are separate, i.e., each $Q_i$ has a different pseudonym $p_i$. The chosen $n_0$ platforms run a a distributed random number generation protocol (e.g., RandHound~\cite{syta2017randhound}) to establish an initial $seed_0$ that is used to bootstrap seed generation for the following rounds. The platforms also produce a joint proof $\pi_0$ that the seed was generated correctly (e.g., the seed signed by all participants). The creator constructs a genesis block 
$$\textsf{Block}_0 = (pk_1, Q_1, pk_2, Q_2, ..., h_e, seed_0, \pi_0, id)$$ 

that includes public keys $pk_i$ and the signed quote structures $Q_i$ for each initial member, a hash of the enclave code $h_e$, the initial seed $seed_0$ and proof $\pi_0$, and a hash $id$ over all elements that serves as the chain identifier. The creator publishes the block and sends the IAS access credential $c_a$ to the attested enclaves that seal it. 

\paragraph{Enrollment.}
After initialization, the system proceeds in rounds that are explained in Section~\ref{sec:operation}. New participants can request enrollment to the system on any round. The joining platform installs the enclave code defined by $h_e$, creates a key pair, seals the private part $sk_n$, exports the public part $pk_n$ and contacts one of the current members, e.g., by broadcasting to the peer-to-peer network.
    
The current member performs remote attestation on the new platform using $h_e$ as the reference. During attestation, the enclave of the new platform supplies a hash $h(id||pk_n||r||h_b)$ as its \textsf{USERDATA}, where $id$ is the chain identifier, $r$ the round number and $h_b$ the hash of the latest block (to bind the enrollment to a specific branch). If the attestation is successful, the existing member obtains a signed \textsf{QUOTE} structure $Q_n$ from IAS, including an attestation pseudonym $p_n$. It verifies that the pseudonym $p_n$ does not appear in any of the previously enrolled platforms in the chain (recall that each $Q_i$ is saved to the ledger). The verifier sends $c_a$ to the attested enclave and constructs an enrollment message
$$\textsf{Enroll}_n = (Q_n, pk_n, r, h_b)$$ 

and broadcasts it to the network. Once the enrollment message is included to a new block, a new identity is established.
        
\paragraph{Re-enrollment.} 
If an enrolled identity does not participate in the system (by sending confirmation messages) for sufficiently long, it will be excluded from leader candidate and endorser selection. In such cases, the platform can perform the enrollment protocol again. In re-enrollment, a chosen verifier checks that the IAS service returns the same pseudonym $p_i$ that was used for this identity (public key $pk_i$) during enrollment. If this is the case, the verifier can create and broadcast a new enrollment message with a flag that indicates re-enrollment. Once such re-enrollment is recorded to a new block, the platform is included to leader candidate and endorser selection again.

\subsection{Mining Identities}
\label{sec:difer:identities}

Our second approach is to ``mine'' identities, that is, reward successful block creation with new identities. A possible strawman solution would be to reward block creation with new coins and each owned coin directly corresponds to one or more identities in the system. As creation of new coins is recorded to the ledger, on each round the owner of the oldest identity can be chosen as the miner. However, this strawman has one major limitation: different coins of same denomination would have different market values. If a coin is old and soon eligible for block creation, its market value is arguably higher than that of a new coin. The fact that the units of the same denomination all have the same value is an important property of any monetary system.

To avoid this problem, we \emph{decouple} coins and identities. Every mining operation creates the normal reward, such as new stake, and additionally an \emph{identity reward}. Sufficiently many ($N_{r} \geq 1$) identity rewards can be used to enroll a new identity to the system. By adjusting the value of $N_r$ it is possible to control the rate of new identities entering the system. The identity rewards can be used in two ways: the block creator can enroll a new identity for himself or he can sell them to a new user that wants to participate in the system.

\paragraph{Initialization.}
The initial distribution of identities can be established using a preliminary PoW phase. 
The initial distribution of identities is the series of public keys $pk_0, pk_1, ...$ from the sequence of blocks $\textsf{InitBlock}_0 = (pk_0, pow_0), \textsf{InitBlock}_1 = (pk_1, pow_1), ...$

Once the initial $n_0$ identities have been created, the participants controlling these identities run a distributed randomness protocol, such as RandHound~\cite{syta2017randhound}, to create the initial random $seed_0$ and the matching proof $\pi_0$ of the correctness of this protocol run that are both attached to the last initial block:
$\textsf{InitBlock}_{n_0} = (pk_{n_0}, pow_{n_0}, seed_0, \pi_0)$. The hash of this block is used as an identifier $id$ for the new chain.


\paragraph{Enrollment.}
Once an identity $pk_m$ has created $N_r$ blocks, it creates a new key pair $(pk_n, sk_n)$ that it uses for enrollment. The enrollment message 
$$\textsf{Enroll} = (h_1, h_2, ..., h_p, pk_n, sig_m)$$ 

contains a set of hashes $\{h_i\}$ that refer to the $N_r$ previously created blocks by $pk_m$, the public key of the new identity $pk_n$, and a signature $sig_m$ over these elements using the private part of $pk_m$. The participant broadcasts the enrollment message, and once it is included to a new block (see Section~\ref{sec:operation}), a new entity exists in the system.
Validity of the enrollment message requires that (1) the set of hashes $h_i$ refer to previous $N_r$ valid blocks, (2) the previous blocks have not been used to create a new identity already, (3) all the referred previous blocks have been created by the same identity $pk_m$, and (4) the enrollment message signature $sig_m$ is correct.

The same mechanism can also be used to allow new participants to join the system. The owner of the identity rewards can sell them to another participant by including a public key received from the buyer to the enrollment message. The payment from the buyer to the seller can be realized by using fiat money or smart contracts. The buyer should release the money only once he sees the correct enrollment message in the chain in a block that has been extended with $d$ valid blocks to prevent double selling of identity rewards.


\section{System Operation}
\label{sec:operation}

Once the initial $n_0$ identities are established, our system proceeds in rounds that have fixed length $t_r$. Next, we explain the system operation on each round $r$.

\subsection{Candidate and Endorser Selection}

In the beginning of each round, every identity tests if it is a leader candidate or endorser. The number of leader candidates $N_c$ and endorsers $N_e$ are both fixed values (e.g., $N_c=5, N_e=100$).
Below, we will informally describe algorithms for candidate and endorser selection. We focus on presentation simplicity; actual implementations can deploy straightforward optimizations like caching previous values. Pseudocode for the algorithms is provided in Appendix~\ref{app:algorithms}.

\vspace{5pt}
\textsf{SelectCandidates()} parses the chain based on two adjustable parameters: activity threshold $T_a$ and $N_c$. An example activity threshold is $T_a=20,000$ rounds that matches one full day of operation. First, the algorithm selects the chain branch to use (see \textsf{SelectBranch} below). Then, it parses the selected branch starting from the newest block till the $T_a$ oldest block. All the identities with recorded confirmation messages in this period are marked as active. Next, the algorithm finds the age of active identities and it marks an identity as inactive, when it has been the oldest for previous $N_C$ rounds (to exclude it from selection if it is not responding). Finally, it sorts this list by age and returns the $N_c$ oldest identities and their ages. 

\vspace{5pt}
\textsf{SelectEndorsers()} computes a list of recently active identities as explained above. If an identity was created less than enrollment threshold $T_e$ rounds ago (e.g., $T_e=100$) it will be excluded from selection to prevent grinding. The algorithm selects $N_e$ identities using standard simple random sampling (with replacement), where identities are sorted based on their public key binary. Random sampling uses $seed_{r-d}$ from the stable part of the chain.  

\subsection{Endorsement Protocol}

\begin{figure}[t]
  \centering
  \includegraphics[width=\linewidth]{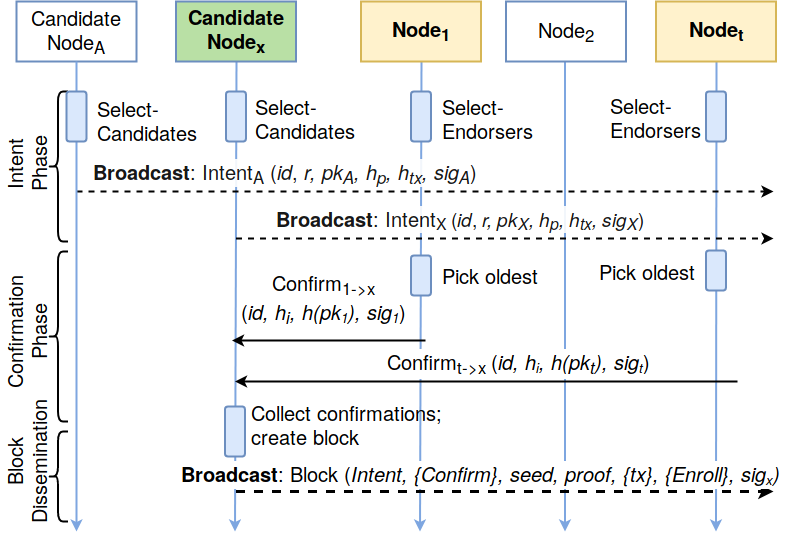}
  \caption{Endorsement protocol. 
  Each leader candidate broadcasts \textsf{Intent} messages and endorsers reply with \textsf{Confirm} messages. A node that receives the required quorum of confirmations becomes an eligible leader for that round and can broadcast a new block.
  }
  \label{fig:endorsement}
  \vspace{-5pt}
\end{figure}

Once the leader candidates and endorsers have been selected, each candidate runs an interactive protocol, shown in Figure~\ref{fig:endorsement}, with the endorsers. The protocol consists of three fixed-length phases.

\paragraph{Intent phase.} 
Each leader candidate $c$ broadcasts 
$$\textsf{Intent} = (id, pk_c, r, h_p, h_{tx}, sig_c)$$ 

message that contains the chain identifier $id$, the candidate's identity $pk_c$, the current round number $r$, the hash of the previous block $h_p$, hash of the transactions $h_{tx}$ the candidate proposes to include in the next block, and the candidate's signature $sig_c$ over these elements. In case multiple chain branches, the candidate uses \textsf{SelectBranch}, described below, to choose which branch to extend.

\paragraph{Confirmation phase.} 
Each endorser $e$ verifies all \textsf{Intent} messages received during the intent phase by checking that the sender is a valid leader candidate. Among the valid \textsf{Intent} messages, the endorser selects the oldest candidate and sends to it $$\textsf{Confirm}_{e \rightarrow c} = (id, h_{i}, h(pk_v), sig_e)$$ 

message that indicates that endorser $e$ has confirmed candidate $c$. This message contains the chain identifier $id$, hash of the intent message $h_{i}$, the endorser identity $h(pk_e)$, and a signature $sig_e$ over the previous elements.
In case multiple candidates have the same age (i.e., they were enrolled in the same block), we choose the oldest candidate in the order their enrollment messages appear in the block. If an endorser receives intent messages that refer to more than one chain branches, the endorser picks the branch to confirm using \textsf{SelectBranch}.
    
\paragraph{Block dissemination phase.}
If candidate $c$ receives at least $q$ \textsf{Confirm} messages, it is chosen as the leader to create a new block. The leader creates a new random seed $seed_r$ and the matching proof $\pi_r$ using the previous seed: $\{seed_{r}, \pi_{r}\} \leftarrow VRF(sk_{m}, seed_{r-1})$. (The used VRF should be such that given a random input, the output should be random, even when the keys are generated by the adversary \cite{nsec5}.) After that, the leader creates and broadcasts a new 
$$\textsf{Block}_r = (\textsf{Intent}, \allowbreak \{\textsf{Confirm}\}, \{tx\},
\{\textsf{Enroll}\}, seed_r, \pi_r, sig_c)$$ 

that contains his intent, the received confirmations, new transactions $\{tx\}$, enrollment messages of new identities, the new seed $seed_r$, the matching proof $\pi_r$, and a signature $sig_c$ over these elements.

\subsection{Chain Validation}

Chain validity is verified using the following algorithms.

\vspace{5pt}
\textsf{SelectBranch()} selects the valid branch among multiple choices. First, it verifies the correctness of each branch using \textsf{VerifyBranch}. Then, it computes a length for each of the branches which is defined by the number of rounds with missing blocks and selects the longest branch. If more than one branch has the same length, it chooses the branch with the older leader at the point of divergence (separate leaders with the same age are chosen based on their enrollment order). If both blocks in the point of divergence were created by the same leader, the  branch is chosen based on the binary of the divergent blocks.

\vspace{5pt}
\textsf{VerifyBranch()} checks that a given chain branch is correctly constructed. It traverses the chain and checks that each block contains a correct hash of the previous block. For each block, it verifies the VRF proof of the random seed. All new identities must have correct \textsf{Enroll} messages (with valid attestation evidence or identity rewards). The algorithm verifies that the miner of each block was a candidate on that round (\textsf{SelectCandidates}), the block contains $q$ confirmations, the confirmation messages contain the hash of the \textsf{Intent} message included to the block, the set of included transaction match $h_{tx}$ from \textsf{Intent}, and the endorsers were eligible on that round (\textsf{SelectEndorsers}).


\section{Security Analysis}
\label{sec:analysis}

In this section, we analyze the security of our solution. For our analysis we use the definition of \emph{stability} from Bonneau et al.~\cite{Bonneau-SP15} with minor adaptation. We say that a consensus scheme is stable if it provides:

\begin{packed_itemize}
	\item \emph{Eventual consensus.} At any time, all honest nodes agree upon a \emph{prefix} of what will eventually become the valid blockchain. 
    
    \item \emph{Exponential convergence.} The probability of a fork at depth $d$ in the chain is $O(2^{-d})$. That is, after a transaction is added to a block that is extended with a small number of valid blocks, the transaction is permanently part of the chain.
    
    \item \emph{Liveness.} New blocks continue to be added and valid transactions included in the blockchain within a reasonable amount of time.

	\item \emph{Correctness.} All the blocks in the prefix of the eventually valid chain will only include valid transactions.

    \item \emph{Fairness.} On expectation, a consensus participants with fraction $\alpha$ of all stake will create fraction $\alpha$ of all blocks, and collect a similar fraction of block creation rewards.
\end{packed_itemize}

\subsection{Consensus and Convergence}
\label{sec:security:consensus}

We first consider the benign case where no participant intentionally manipulates the random seed. After that, we consider the more complicated case where the attacker manipulates the seed.

To examine the different possible cases in leader endorsement, consider an example where the three oldest leader candidates are $A$, $B$ and $C$. The endorser committee is sampled based on $seed_{r-d}$. When $seed_{r-d}$ is unbiased and when the identities that take part in the sampling have been fixed before $seed_{r-d}$ is known (as is the case in our solution), on the average $\alpha$ of the sampled endorsers are adversary-controlled. Fraction $\beta$ of the endorsers may not receive \textsf{Intent} sent by the oldest candidate $A$. The remaining fraction $1-\alpha-\beta$ of endorsers who received all messages, confirm $A$ as the leader. Those endorsers that did not receive all messages may confirm another candidate ($B$ or $C$). The adversary-controlled endorsers may confirm more than one candidate ($A$ and $B$), although such equivocation leaves evidence that can be easily used to penalize malicious identities (see Section~\ref{sec:discussion} for discussion).


In a rare case, at least $q$ endorsers are sampled from the fraction $\alpha + \beta$ of active identities. In such case, also the second-oldest candidate $B$ may receive the required confirmations, causing two eligible leaders ($A$ and $B$) and a fork in the chain. We denote the probability of such \emph{benign fork sampling} as \textsf{Pr(BFS)}. Assuming sufficiently many active identities $n_a$, it can be computed as:
$$ \textsf{Pr(BFS)} = \sum\limits_{i=0}^{N_e-q} \bigg( \binom{N_e}{q+i} (\alpha + \beta)^{q+i} (1-\alpha-\beta)^{N_e-q-i} \bigg). $$

For example, when $\alpha = 0.33$, $\beta = 0.05$, $N_e=100$ and $q=54$, then \textsf{Pr(BFS)} = 0.0008. That is, such sampling would take place, on the average, every 1200 rounds. Extending both forked branches requires another similar sampling. As the probability of consecutive sampling decreases exponentially and the probability of three consecutive samplings is already very low ($5.78 \times 10^{-10}$). For such parameter values we consider the maximum depth of forks $d=3$ in the absence of seed manipulation. 

Next, we consider the adversarial case where the attacker intentionally manipulates $seed_r$ to bias endorser selection. Recall that we use VRFs to update the seed for each new block. If the adversary controls more than one oldest leader candidates, it may choose which one of these identities it uses to create the block and update the seed. This gives the adversary more than one seed alternatives to choose from. If the adversary similarly controls more than one oldest leader candidate on the next round, he can again choose which candidate to use to update the seed. Such process allows the adversary to build a ``seed prediction tree'' which expansion factor is the number of controlled oldest candidates on each round and which depth is the number of successive rounds where the adversary controls more than one oldest candidate. Since identity enrollment is permissionless and open, the adversary may control multiple oldest candidates on several successive rounds and build a large seed prediction tree.

Assume an adversary that on round $r$ builds a seed prediction tree of depth $d_t$ and with $2^{80}$ leafs. We consider that building a tree larger than that is infeasible, as the tree needs to be constructed online without pre-computation. This tree allows the adversary to pick the seed update schedule on round $r$ that will give the most beneficial endorser sampling sequence starting from round $r+d_t$ out of the $2^{80}$ predicted options. Recall from our above analysis that, given our example parameter values, benign fork sampling probability $\textsf{Pr(BFS)}=0.0008$. The probability of finding such sampling on, for example, $d=12$ successive rounds reduces exponentially and becomes very low ($6.87 \times 10^{-32}$). With the above seed prediction tree, the adversary has $2^{80}$ attempts to find such a sequence of samplings. We call the probability that the adversary finds such \emph{adversarial fork sampling} as \textsf{Pr(AFS)} $= 6.87 \times 10^{-32} \times 2^{80} = 8.3 \times 10^{-14}$. Thus, setting the maximum depth of forks to $d=12$ prevents such attacks. 

Because identity enrollment is open, such seed prediction attacks cannot be prevented altogether. However, in Section~\ref{sec:discussion} we discuss how such attacks can be made difficult to realize in practice by using multiple identity queues and forcing the adversary to plan the attack years before its execution.

\begin{figure}[t]
  \centering
  \begin{subfigure}[b]{0.8\linewidth}
    \includegraphics[width=\linewidth]{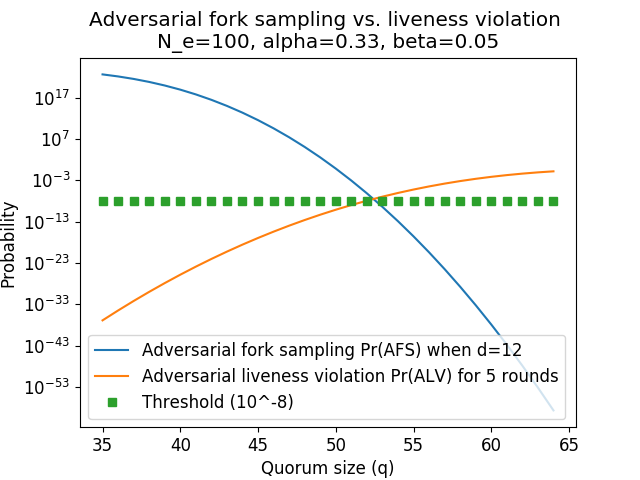}
    \caption{For parameter values $N_e=100$, $\alpha=0.33$, $\beta=0.05$ quorum $q=54$ prevents forks at depth $d=12$ without compromising liveness.}
    \label{fig:Ne100}
  \end{subfigure}
  \begin{subfigure}[b]{0.8\linewidth}
    \includegraphics[width=\linewidth]{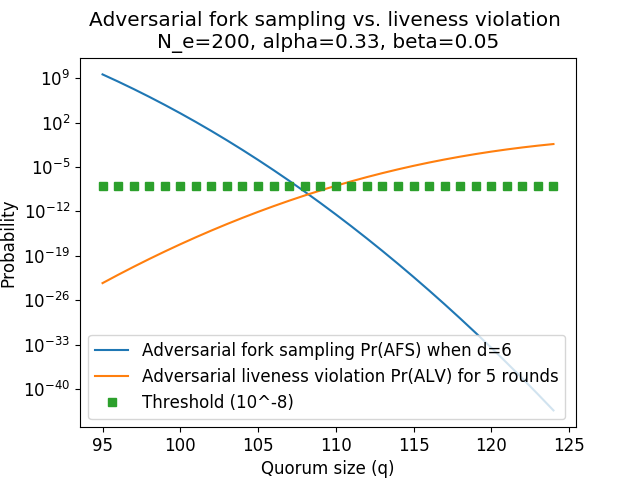}
    \caption{For parameter values $N_e=200$, $\alpha=0.33$, $\beta=0.05$ quorum $q=108$ prevents forks at depth $d=6$ without compromising liveness.}
    \label{fig:Ne200}
  \end{subfigure}
  \caption{The quorum size $q$ represents a trade-off between security and liveness. As $q$ increases, the adversarial fork sampling probability \textsf{Pr(AFS)} reduces and the adversarial liveness violation probability \textsf{Pr(ALV)} increases.}
  \label{fig:forks-vs-liveness}
  \vspace{-10pt}
\end{figure}

Increasing the quorum size $q$ reduces the probability of forks at depth $d$, but weakens liveness guarantees as explained later. Figure~\ref{fig:Ne100} shows that quorum value $q=54$ provides a good balance of security and liveness when $N_e=100$.

Increasing number of endorsers $N_e$ also reduces $d$. As shown in Figure~\ref{fig:Ne200}, when $N_e=200$ endorsers are used, forks can be reduced to $d=6$ rounds without compromising liveness. The main drawback of larger $N_e$ is that such solution requires more communication on each round. We discuss system performance and communication complexity in more detail in Section~\ref{app:evaluation}.

In Appendix~\ref{app:extra-analysis} we extend this analysis to consider different parameter values, including stronger adversaries (e.g., $\alpha=0.4$), better connectivity (e.g., $\beta=0.01$), and larger endorser committees (e.g., $N_e=400$).

\subsection{Liveness} 

Block creation requires that at least one of the leader candidates receives $q$ confirmations. We first consider the benign case where all endorsers confirm the oldest \textsf{Intent} they receive. We denote the probability that more than $N_e-q$ endorsers will be sampled from the fraction of $\beta$ identities that did not receive the \textsf{Intent} message as \emph{benign liveness violation} \textsf{Pr(BLV)} and compute it as:
$$ \textsf{Pr(BLV)} = \sum\limits_{i=0}^{q} \bigg( \binom{N_e}{N_e-q+i} \beta^{N_e-q+i} (1-\beta)^{q-i} \bigg). $$

Given the previous example parameters, this probability is negligible ($6.96 \times 10^{-33}$).
Next, we consider the case where the adversary reduces the probability of successful block creation by intentionally not sending \textsf{Confirm} messages to targeted leader candidates. Such \emph{adversarial liveness violation} probability \textsf{Pr(ALV)} can be computed as:
$$ \textsf{Pr(ALV)} = \sum\limits_{i=0}^{q} \bigg( \binom{N_e}{N_e-q+i} (\alpha+\beta)^{N_e-q+i} (1-\alpha-\beta)^{q-i} \bigg). $$

Given the previous example parameters, the adversary can prevent mining with probability 0.062, that is, on the average every 16th round. 
The probability to prevent mining on five successive rounds is $9.37 \times 10^{-7}$ (see Figure~\ref{fig:Ne100}).
If the adversary continues this strategy longer than the activity period $T_a$, its identities will be considered inactive and they can no longer reduce the mining probability for other participants. 

As in any leader-based blockchain consensus scheme, the chosen leader can exclude transactions from targeted users, and therefore no such scheme can provide an absolute guarantee that a new transaction is included to the next block. In our approach, this problem is somewhat exacerbated. Since the adversary may control block creation on multiple successive rounds, it may prevent inclusion of specific transactions for a longer time. In this regard, our solution provides a weaker liveness guarantee than schemes based on random leader selection. While we cannot prevent such denial-of-service attacks altogether, in Section~\ref{sec:discussion} we discuss how such attack can be made difficult to realize in practice by using multiple identity queues.

\subsection{Correctness} 

Regarding transaction correctness, similar to any other leader-based consensus scheme, the chosen leader can include invalid transactions to the published block. Users can detect and ignore falsely formatted transactions. Transactions that appear valid in the current branch but contradict transactions in another branch (e.g., double spending) can be detected by waiting $d$ rounds. Thus, all transactions in the chain prefix up to $\textsf{Block}_{r-d}$ are either valid or ignored.

\subsection{Fairness} 

The adversary can attempt to violate fairness in few ways. The first approach is that the adversary does not include \textsf{Enroll} messages from the targeted victim participant to its blocks. This approach can delay enrollment of a new identity by a few rounds, but not prevent it, and thus such an approach does not violate fairness in the long term. The second approach is that the adversary does not include \textsf{Confirm} messages from the victim to its blocks and after $T_a$ rounds the victim is excluded from miner candidate selection and has to re-enroll. Such \emph{adversarial exclusion} probability can be computed as
$$\textsf{Pr(AE)} = (1-N_e/n_a)^{T_a(1-\alpha)}.$$ 

Assuming $n_a=10,000$ active participants and our example parameters, the adversarial exclusion probability is negligible ($3.25 \times 10^{-59}$). If the size of the system increases to $n_a=100,000$ the adversarial exclusion probability is still low ($1.5 \times 10^{-6}$). If the grows larger than that, the value of $T_a$ may have to be increased to ensure that active identities are not excluded from selection.


\subsection{SGX Considerations}

The adversary may attempt to enroll non-SGX platforms, but such false enrollment would fail, as the IAS will not return a signed \textsf{QUOTE} needed for enrollment. Enrolling the same SGX platform multiples times would fail as well, because the IAS would return the same pseudonym $p_n$ that is already recorded for another identity in the chain. The third alternative is to enroll the same SGX platform to multiple chains and try to reuse enrollment from one chain to another. Because the \textsf{QUOTE} contains the chain identifier $id$, this approach would not work either. 

The adversary does not gain advantage (more identities or selection bias) by breaking into her own SGX processors. Besides attestation, we only use enclaves for the protection of the IAS credential and leakage of this credential does not allow the adversary to create additional identities. A malicious chain creator could initialize an invalid chain, where all members are not SGX processors. However, any legitimate participant can detect this due to missing \textsf{QUOTEs} in the genesis block and neglect the chain.

In case the attestation service (IAS) is temporarily unavailable, new identities cannot be enrolled during its downtime. However, the system can produce new blocks and thus process incoming transactions normally. Therefore, the centralized IAS is not critical for liveness.

\subsection{Privacy Considerations} 

Since block creation is based on long-term identities, correlation of block creation events by the same participant becomes trivial. This is a limitation of our approach compared consensus systems where participants pick new identities for every round. However, we emphasize that the identities used for transactions can be completely separate from those used for consensus and block creation. For example, transactions can be based on changeable pseudonyms or cryptographic commitments that hide user identities and transaction values~\cite{maxwell2015confidential,mimblewimble}.


\section{Performance Evaluation}
\label{app:evaluation}

In this section we explain the experiments we performed in order to estimate suitable round duration $t_r$, transaction latency and throughput.

\subsection{Experimental Setup}
We built a globally-distributed peer-to-peer network using Amazon's AWS infrastructure. We instantiated nodes in Frankfurt, London, Singapore, Mumbai and Oregon. We used the EC2 compute services with nodes ranging from t2.micro (single vCPU with 1 GB RAM) to m4.2xlarge (8 vCPUs and 16 GB RAM). The node software was written in Java and run on Ubuntu/Linux OS. 
To simulate the worst case scenario, we ensured that the leader candidate was never located in the same data center as any of the endorsers. To simulate global distribution of participants, we enforced that messages have to travel through at least $x$ different nodes ($x$ being 0, 2 and 6) before reaching their destination. We set the \textsf{Intent} and \textsf{Confirm} message sizes to 1 KB (although actual messages are smaller). For blocks we tested for three sizes: 500KB, 1MB and 2MB.

\paragraph{Network optimization}
During testing we observed that majority of the block dissemination delay came from the initial block transmission by the leader candidate, due to a high out-degree and due to multiple hops across geographically distant locations. To address these issues, we implemented a networking structure 
where we selected some nodes within a cluster of geographically close nodes to serve as \emph{top-level nodes}, i.e., nodes that are directly connected to by leaders when broadcasting the block. These top nodes have a large out-degree to mid-level nodes within the same geographical area. This optimization led to a significant reduction in block dissemination latency. (We did not see the need to use the same approach in the intent phase, as it did not lead to any noticeable improvement.)
We emphasize that the top and middle nodes are not different from other network nodes. Any node could be chosen as a top or middle node and messages may be broadcast to multiple top nodes within a cluster. In a large deployment, the top-level nodes may be chosen by reliability and performance metrics similar to the Tor network. 



\subsection{Results}

\begin{figure}[t]
  \centering
  \begin{subfigure}[b]{0.85\linewidth}
  	\includegraphics[width=\linewidth]{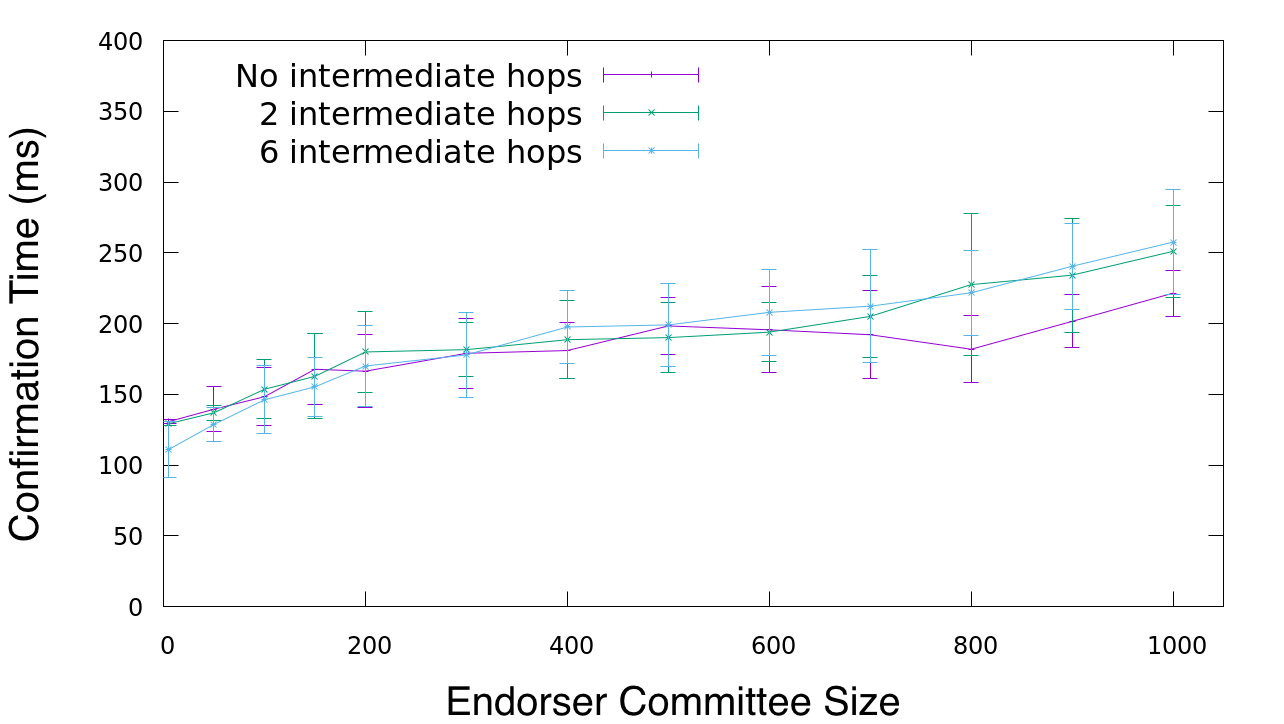}
    \caption{Intent and confirmation time.}
	\label{fig:perf_intent}
  \end{subfigure}
  \begin{subfigure}[b]{0.85\linewidth}
  	\includegraphics[width=\linewidth]{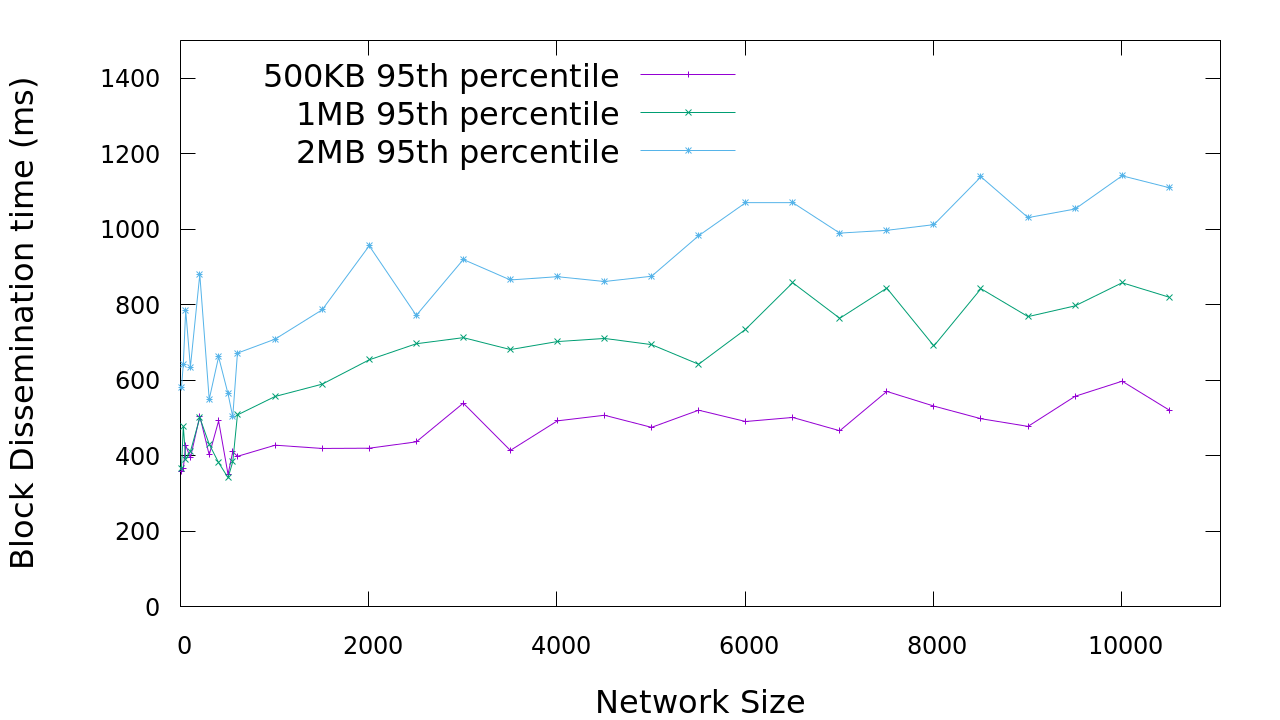}
    \caption{Block dissemination time.}
	\label{fig:perf_block}
  \end{subfigure}
  \caption{Experimental results for message delivery times in our test setup, a globally-distributed peer-to-peer network using Amazon's AWS infrastructure.}
  \label{fig:perf}
  \vspace{-10pt}
\end{figure}

We measured message delivery times for various system and block size. Figure~\ref{fig:perf} summarizes the results of our experiments. 

In Figure~\ref{fig:perf_intent} we plot the time required for leader selection (combined \textsf{Intent} message delivery and \textsf{Confirm} message reception). This time grows from $130$ ms for small endorser committee size $N_e=5$ to $257$ ms for large committee size $N_e=1000$. We conclude that setting the combined duration of these two phases to one second is sufficient in a network environment like ours. 

Figure~\ref{fig:perf_block} shows the time required for block dissemination (95th percentile) that grows from $357$ ms for a system size of $n_a=10$ active nodes to $1.1$ seconds for a system size of $n_a=10,000$ active nodes. We conclude that setting the duration of block dissemination phase to 4 seconds is sufficient for our network. The above two values give us a round duration of $t_r=5$ seconds.

\paragraph{Throughput and latency.}
Given such example round duration $t_r=5$ seconds, we can now estimate system throughput $tp$ for or solution as follows:
$$ tp = \frac{\frac{1}{t_r} \times (B - H - (N_e \times S_C) - (n_a \times S_E))}{T}, $$

where $H$ is invariant block header (280 bytes), $S_C$ is the size of \textsf{Confirm} message (416 bytes), $B$ is the used block size, $T$ is the transaction size, and $S_E$ is the size of the \textsf{Enroll} message. 
Assuming $B=2$ MB and $T=250$ bytes, similar to Bitcoin~\cite{blocksize}, $N_e=100$ endorsers and few enrolments per round (owing to fast rounds), 99\%+ of the block is left for the transactions and the system throughput is approximately 1500 transaction per second.
Transaction latency is one minute (when $d=12$) or 30 seconds ($d=6$).

\section{Discussion}
\label{sec:discussion}


\paragraph{Improved latency and liveness.}
An adversary that controls multiple oldest identities can predict seed evolution which enables deeper forks and thus higher latency. Although seed prediction cannot be prevented completely, it can be made difficult to realize in practice. 
One possible defensive approach is to use \emph{multiple queues}. Recall that in our approach all identities are essentially placed into one queue in the order of their age and the oldest identities are picked as leader candidates in round robin. Instead of using a single queue, identities could be placed into multiple queues. For example, in a system that has been operational for five years and each year equally many identities were enrolled, a separate queue could be established for each enrollment year. The leader selection could happen such that the oldest identity is picked from each queue in turn. To perform successful seed prediction, the adversary would now have to plan the attack years ahead, so that he controls identities in all queues at correct places. The same applies for targeted liveness attacks. To prevent processing of victim's transactions, the adversary would have to plan his enrollment schedule years before the attack takes place.

Besides improved security, multiple queues could be used also as a performance enhancing mechanism. Similar to \emph{sharding}, each queue could process a separate set of incoming transactions in parallel which could increase the system's overall throughput. We consider multiple identity queues an interesting direction for future work.

Predictable leader selection can make denial-of-service attacks easier. For example, the adversary can prepare the attack in advance and launch it when the victim becomes leader candidate. Interestingly, such predictability can also help participants in avoiding DoS attacks. Participants can obtain multiple IP addresses and switch to using a different IP before their identity becomes the leader candidate. Similar defensive approaches are harder to realize in systems with randomized leader selection. 

\paragraph{Bootstrapping from other infrastructures.} 
Similar to Intel SGX, reliable identities could be bootstrapped also from other infrastructures. For example, mobile phone operators, credit card companies or passport issuers could take the role of IAS and provide an interface that allows their customers to enroll new identities in a controlled manner, such as one identity per person or mobile phone subscription. Another attestation infrastructure that could be leveraged is TrustZone smartphones~\cite{trustzone}. Also new and emerging secure processor architectures that are designed specifically for distributed ledger technologies~\cite{gradient} could be used to create identities for our solution. Recent efforts to standardize EPID provisioning and attestation across manufacturers~\cite{epid_standard,EPIDstandard} could provide a vendor independent way of bootstrapping identities. Identities could also be bootstrapped from multiple sources (say, credit card number and TrustZone smartphone and registered phone number) to provide the right security and usability for the particular use case. We focus on SGX identities, as the IAS service enables deployment of our solution today without any changes to existing infrastructures.

\paragraph{Expanding the role of endorsers.}
In our solution, the endorsers confirm the oldest leader candidate they observe, regardless of the content of the block that the candidate proposes to create. The role of the endorsers could be expanded to examine the proposed block, e.g., for validity of proposed transactions. Such optimizations could allow the endorsers to ignore leaders that would extend the chain with invalid blocks and pick a different leader candidate instead.

\paragraph{Penalizing malicious behavior.} 
In most permissionless consensus schemes identities can be easily changed. For example, a Bitcoin miner can use a different public key every time he starts mining for a new block. In our approach identities cannot be changed after the initial enrollment, as they are recorded to the blockchain. One advantage of long-lived identities is that penalizing malicious behavior becomes possible. For example, if an endorser confirms multiple intents on the same round, any entity that observes this can broadcast the conflicting and signed confirmation messages and the next miner can include them to a new block as evidence of cheating which could result in automated elimination of the malicious identity from the system. Thus, participants have an incentive to avoid misbehavior. 

\paragraph{Economic aspects.} 
In case of SGX identities, participants are incentivized to buy the cheapest processors that enable enrollment. If processors have significantly different value, this could raise questions about the fairness~\cite{bano-2017}. 
We argue that Intel, or similar manufacturer, is unlikely to sell unused but outdated products in mass-scale and purchasing cheap second-hand processors may not provide an advantage, because those CPUs may have already been enrolled. Also, enrollment of very old CPUs can be prevented. In SGX, the attestation group signature does not identify the individual CPU but it does reveal the manufacturing batch.

\section{Related Work}
\label{sec:related-work}

In Section~\ref{sec:background} we  outlined the limitations of several related solutions. In this section we review additional related work. For a general comparison and classification of blockchain consensus, we refer the reader to~\cite{bano-2017}.

\paragraph{Other Proof-of-Stake schemes.} Ouroboros Praos~\cite{praos} is another PoS scheme that leverages VRFs for new random value generation on each round, similar to Algorand~\cite{algorand}. The main limitation of this approach is that such randomness can be biased and thus the solution does not provide fairness.

RapidChain~\cite{rapidchain} samples a reference committee from all consensus participants. The reference committee is then responsible for running a distributed randomness generation protocol in the beginning of each epoch to create new randomness for that epoch. The randomness protocol is based on verifiable secret sharing (VSS). The main limitations of this approach is that the reference committee becomes an obvious target for attacks and the distributed random generation protocol is expensive.

DFINITY~\cite{dfinity} introduces a novel decentralized an random beacon that leverages BLS threshold signatures for periodic unbiased random values generation. This scheme requires a setup phase during which an expensive distributed key generation (DKG) protocol is run. Once this is done, new random values can be derived by collecting signature shares from sufficiently many participants. In this approach, the per round or per epoch randomness generation has low communication complexity, but the main cost is the expensive DKG protocol in the setup phase that needs to be repeated when new participants join or leave the system.

\paragraph{Other TEE solutions} Proof of Luck (PoL)~\cite{luck} is SGX-based solution that has the same basic idea and the same main limitations as PoET (recall Section~\ref{sec:background}). 

PoTS~\cite{PoTS} is another PoS solution that uses SGX and is designed to tolerate compromised TEEs that control up to 50\% stake. One drawback of this approach is that compromising a small number of high-stake TEEs it may be possible to compromise of the entire system (due to concentration of stake to few rich individuals). Moreover, the approach does not provide fairness. Finally, PoTS requires TEEs, while our solution works also without them.

Resource Efficient Mining (REM)~\cite{REM} replaces the hash computation of PoW with attested enclave computation. This approach allows more useful usage of energy, but it does not eliminate the need for massive collective computation. Our approach requires no solving of computationally intensive puzzles and thus it saves significant amounts of energy compared to PoW-based solutions.

\paragraph{Coin aging.} PPCoin~\cite{king2012ppcoin} introduced the idea that each coin has an associated age and leader selection is based on hashing procedure where the target difficulty is coin-specific and lower for older coins. However, the suggested scheme is vulnerable to a simple attack where the adversary waits so that he owns enough old coins and then creates a deep fork for double spending. The authors suggest that such attacks could be addressed with a central time-stamping mechanism which is a circular argument for a decentralized and permissionless blockchain consensus scheme. Additionally, the leader selection is not fair, because selection can be manipulated with simple grinding strategies.


\section{Conclusion}
\label{sec:conclusion}

In this paper we have explored an alternative idea for blockchain consensus --- selecting consensus leader candidates \emph{deterministically} instead of the common \emph{random} selection approach and complementing such selection with a simple interactive endorsement protocol. The main benefits of our solution are simplicity and fairness. As our analysis shows, the latter is especially important in systems where block creation is rewarded with new stake which is a common practice is permissionless blockchains. Although deterministic selection has also its own limitations (weaker DoS resilience), this work shows that it provides a viable and previously unexplored alternative to random selection.


\bibliographystyle{plain}
\bibliography{references}

\begin{thebibliography}{10}

\bibitem{sawtooth}
Sawtooth lake documentation, 2016.
\newblock \url{https://intelledger.github.io/introduction.html}.

\bibitem{sgx_ias_api}
Attestation service for intel® software guard extensions (intel sgx): Api
  documentation, 2017.
\newblock
  \url{https://software.intel.com/sites/default/files/managed/7e/3b/ias-api-spec.pdf}.

\bibitem{sgxwebpage}
Intel sgx homepage, 2017.
\newblock \url{https://software.intel.com/en-us/sgx}.

\bibitem{gradient}
Introducing gradient, August 2018.
\newblock
  \url{https://medium.com/gradient-tech/introducing-gradient-715e11be685b}.

\bibitem{PoTS}
Sébastien Andreina, Jens-Matthias Bohli, Ghassan~O. Karame, Wenting Li, and
  Giorgia~Azzurra Marson.
\newblock Pots - a secure proof of tee-stake for permissionless blockchains.
\newblock Cryptology ePrint Archive, Report 2018/1135, 2018.
\newblock \url{https://eprint.iacr.org/2018/1135}.

\bibitem{trustzone}
ARM.
\newblock Trustzone.
\newblock \url{https://www.arm.com/products/security-on-arm/trustzone}.

\bibitem{bano-2017}
Shehar Bano, Alberto Sonnino, Mustafa Al-Bassam, Sarah Azouvi, Patrick McCorry,
  Sarah Meiklejohn, and George Danezis.
\newblock Consensus in the age of blockchains.
\newblock {\em arXiv preprint arXiv:1711.03936}, 2017.

\bibitem{blocksize}
StackExchange Bitcoin.
\newblock What is the average size of a bitcoin transaction?
\newblock
  \url{https://bitcoin.stackexchange.com/questions/31974/what-is-the-average-size-of-a-bitcoin-transaction}.

\bibitem{bitcoin-stats}
BitcoinStats.
\newblock Data propagation: Daily snapshots.
\newblock \url{http://bitcoinstats.com/network/propagation/}.

\bibitem{Bonneau-SP15}
Joseph Bonneau, Andrew Miller, Jeremy Clark, Arvind Narayanan, Joshua~A. Kroll,
  and Edward~W. Felten.
\newblock Sok: Research perspectives and challenges for bitcoin and
  cryptocurrencies.
\newblock In {\em IEEE Symposium on Security and Privacy (S\&P)}, 2015.

\bibitem{sgx_kari_sidechannel}
Ferdinand Brasser, Urs M{\"{u}}ller, Alexandra Dmitrienko, Kari Kostiainen,
  Srdjan Capkun, and Ahmad{-}Reza Sadeghi.
\newblock Software grand exposure: {SGX} cache attacks are practical.
\newblock {\em CoRR}, abs/1702.07521, 2017.

\bibitem{cachinrandomgen}
Christian Cachin, Klaus Kursawe, Frank Petzold, and Victor Shoup.
\newblock Secure and efficient asynchronous broadcast protocols.
\newblock In {\em Advances in Cryptology - {CRYPTO} 2001, 21st Annual
  International Cryptology Conference, Santa Barbara, California, USA, August
  19-23, 2001, Proceedings}, pages 524--541, 2001.

\bibitem{pos_history}
CoinTelegraph.
\newblock The history and evolution of proof of stake.
\newblock
  \url{https://cointelegraph.com/news/the-history-and-evolution-of-proof-of-stake}.

\bibitem{praos}
Bernardo David, Peter Ga{\v{z}}i, Aggelos Kiayias, and Alexander Russell.
\newblock Ouroboros praos: An adaptively-secure, semi-synchronous
  proof-of-stake blockchain.
\newblock In {\em Annual International Conference on the Theory and
  Applications of Cryptographic Techniques}, pages 66--98. Springer, 2018.

\bibitem{Decker-P2P13}
Christian Decker and Roger Wattenhofer.
\newblock {Information Propagation in the Bitcoin Network}.
\newblock In {\em International Conference on Peer-to-Peer Computing (P2P)},
  2013.

\bibitem{digiconomist}
Digiconomist.
\newblock Bitcoin energy consumption index.
\newblock \url{https://digiconomist.net/bitcoin-energy-consumption}.

\bibitem{dwork-1992}
Cynthia Dwork and Moni Naor.
\newblock Pricing via processing or combatting junk mail.
\newblock In {\em Annual International Cryptology Conference}, pages 139--147.
  Springer, 1992.

\bibitem{garay-2015}
Juan~A Garay, Aggelos Kiayias, and Nikos Leonardos.
\newblock The bitcoin backbone protocol: Analysis and applications.
\newblock In {\em EUROCRYPT (2)}, pages 281--310, 2015.

\bibitem{nsec5}
Sharon Goldberg, Moni Naor, Dimitrios Papadopoulos, and Leonid Reyzin.
\newblock {NSEC5} from elliptic curves: Provably preventing {DNSSEC} zone
  enumeration with shorter responses.
\newblock {\em {IACR} Cryptology ePrint Archive}, 2016:83, 2016.

\bibitem{dfinity}
Timo Hanke, Mahnush Movahedi, and Dominic Williams.
\newblock Dfinity technology overview series, consensus system.
\newblock {\em arXiv preprint arXiv:1805.04548}, 2018.

\bibitem{epid_standard}
Intel.
\newblock Atmel and microchip adopt intel identity technology for iot.
\newblock
  \url{http://download.intel.com/newsroom/kits/idf/2015_fall/pdfs/Intel_EPID_Fact_Sheet.pdf}.

\bibitem{EPIDstandard}
Intel.
\newblock A cost-effective foundation for end-to-end iot security, 2017.
\newblock
  \url{https://www.intel.com/content/dam/www/public/us/en/documents/white-papers/intel-epid-white-paper.pdf}.

\bibitem{intel-vulnerability}
Intel.
\newblock Q3 2018 speculative execution side channel update, August 2018.
\newblock
  \url{https://www.intel.com/content/www/us/en/security-center/advisory/intel-sa-00161.html}.

\bibitem{mimblewimble}
Tom~Elvis Jedusor.
\newblock Mimblewimble.
\newblock \url{http://mimblewimble.org/mimblewimble.txt}.

\bibitem{Ouroboros2017}
Aggelos Kiayias, Alexander Russell, Bernardo David, and Roman Oliynykov.
\newblock {Ouroboros: A provably secure proof-of-stake blockchain protocol}.
\newblock {\em Lecture Notes in Computer Science (including subseries Lecture
  Notes in Artificial Intelligence and Lecture Notes in Bioinformatics)}, 10401
  LNCS:357--388, 2017.

\bibitem{peercoin}
Sunny King and Scott Nadal.
\newblock Peercoin whitepaper, 2012.
\newblock \url{https://peercoin.net/whitepaper}.

\bibitem{king2012ppcoin}
Sunny King and Scott Nadal.
\newblock Ppcoin: Peer-to-peer crypto-currency with proof-of-stake.
\newblock {\em self-published paper, August}, 19, 2012.

\bibitem{Lipp2018meltdown}
Moritz Lipp, Michael Schwarz, Daniel Gruss, Thomas Prescher, Werner Haas,
  Stefan Mangard, Paul Kocher, Daniel Genkin, Yuval Yarom, and Mike Hamburg.
\newblock Meltdown.
\newblock {\em ArXiv e-prints}, January 2018.

\bibitem{maxwell2015confidential}
Gregory Maxwell.
\newblock Confidential transactions.
\newblock \url{https://people.xiph.org/\~greg/confidential\_values.txt}, 2015.

\bibitem{algorand}
Silvio Micali.
\newblock {ALGORAND:} the efficient and democratic ledger.
\newblock {\em CoRR}, abs/1607.01341, 2016.

\bibitem{vrf-micali-99}
Silvio Micali, Michael Rabin, and Salil Vadhan.
\newblock Verifiable random functions.
\newblock In {\em Foundations of Computer Science, 1999. 40th Annual Symposium
  on}, pages 120--130. IEEE, 1999.

\bibitem{luck}
Mitar Milutinovic, Warren He, Howard Wu, and Maxinder Kanwal.
\newblock Proof of luck: An efficient blockchain consensus protocol.
\newblock In {\em Proceedings of the 1st Workshop on System Software for
  Trusted Execution}, SysTEX '16, pages 2:1--2:6, New York, NY, USA, 2016. ACM.

\bibitem{bitcoin}
Satoshi Nakamoto.
\newblock Bitcoin: A peer-to-peer electronic cash system.
\newblock May 2009.

\bibitem{sgx_sidechannel}
Michael Schwarz, Samuel Weiser, Daniel Gruss, Cl{\'{e}}mentine Maurice, and
  Stefan Mangard.
\newblock Malware guard extension: Using {SGX} to conceal cache attacks.
\newblock {\em CoRR}, abs/1702.08719, 2017.

\bibitem{syta2017randhound}
Ewa Syta, Philipp Jovanovic, Eleftherios~Kokoris Kogias, Nicolas Gailly, Linus
  Gasser, Ismail Khoffi, Michael~J Fischer, and Bryan Ford.
\newblock Scalable bias-resistant distributed randomness.
\newblock In {\em Security and Privacy (SP), 2017 IEEE Symposium on}, pages
  444--460. Ieee, 2017.

\bibitem{van2018foreshadow}
Jo~Van~Bulck, Frank Piessens, and Raoul Strackx.
\newblock Foreshadow: Extracting the keys to the intel $\{$SGX$\}$ kingdom with
  transient out-of-order execution.
\newblock In {\em 27th $\{$USENIX$\}$ Security Symposium ($\{$USENIX$\}$
  Security 18)}, 2018.

\bibitem{rapidchain}
Mahdi Zamani, Mahnush Movahedi, and Mariana Raykova.
\newblock Rapidchain: Scaling blockchain via full sharding.
\newblock In {\em Proceedings of the 2018 ACM SIGSAC Conference on Computer and
  Communications Security}, pages 931--948. ACM, 2018.

\bibitem{REM}
Fan Zhang, Ittay Eyal, Robert Escriva, Ari Juels, and Robbert van Renesse.
\newblock {REM:} resource-efficient mining for blockchains.
\newblock {\em {IACR} Cryptology ePrint Archive}, 2017:179, 2017.

\end{thebibliography}


\clearpage
\appendix


\section{Pseudocode for Algorithms}
\label{app:algorithms}

In this appendix, we provide pseudocode for algorithms that were described in Section~\ref{sec:operation}. For presentation simplicity, we omit simple optimizations, such as caching, and trivial checks and helper functions.

\begin{algorithm}
\footnotesize
\caption{SelectCandidates}\label{MiningAlgorithmsSelectCandidates}
\begin{algorithmic}[1]
\Procedure{SelectCandidates}{\textit{branch}}
\State $\textit{ActiveSet\{\}} \gets \textsc{SelectActive(}\textit{branch)}$
\State $\textit{SortByAge(ActiveSet)}$
\State $\textit{CandidateSet\{\}} \gets \textit{ActiveSet[0:}N_{c-1}\textit{]}$
\While {\textit{InactiveRounds(CandidateSet[0])} $\geq N_c$}
\State \textit{CandidateSet[0]} $\gets$ \textit{Inactive}
\State \textit{CandidateSet[$N_c$]} $\gets$ \textit{ActiveSet[$N_c$]}
\State \textit{ShiftLeft(CandidateSet,1)}
\EndWhile
\State \Return \textit{CandidateSet}
\EndProcedure
\end{algorithmic}
\end{algorithm}

\begin{algorithm}
\footnotesize
\caption{SelectEndorsers}\label{MiningAlgorithmsSelectEndorsers}
\begin{algorithmic}[1]
\Procedure{SelectEndorsers}{\textit{branch}}
\State \textit{ActiveNodes\{\}} $\gets$ \textsc{SelectActive(\textit{branch})}
\ForAll{\textit{node} $\in$ \textit{ActiveNodes}}
\If{\textit{EnrollmentAge(node) $< T_e$ }}
\State \textit{ActiveNodes $=$ ActiveNodes \textbackslash node}
\EndIf
\EndFor
\State \textit{EndorserSet $\gets$ \textit{RandomSampling($Seed_{r-d}$, ActiveNodes)}}
\State \Return \textit{EndorserSet}
\EndProcedure
\end{algorithmic}
\end{algorithm}

\begin{algorithm}
\footnotesize
\caption{SelectActive}\label{MiningAlgorithmsSelectActive}
\begin{algorithmic}[1]
\Procedure{SelectActive}{\textit{branch}}
\State $\textit{iterBlock} \gets \textit{Top(branch)}$
\State \textit{i} $\gets$ 0
\State $\textit{ActiveSet\{\}} \gets \emptyset$
\While {$\textit{i} <  T_a$}
\State \textit{i $\gets i+1$}
\State $\textit{BlockEndorsers} \gets \textit{GetEndorsers(iterBlock)}$
\ForAll{$\textit{Endorser}\in\mathit{BlockEndorsers}$}
\If {$\textit{Endorser} \notin \textit{ActiveSet}$}
\State $\textit{ActiveSet} \gets \textit{ActiveSet $\cup$ Endorser}$
\EndIf
\EndFor
\State \textit{iterBlock $\gets$ Next(iterBlock)}
\EndWhile
\State \Return \textit{ActiveSet}
\EndProcedure
\end{algorithmic}
\end{algorithm}

\begin{algorithm}
\footnotesize
\caption{SelectBranch}\label{MiningAlgorithmsSelectBranch}
\begin{algorithmic}[1]
\Procedure{SelectBranch}{\textit{Branches\{\}}}
\ForAll{\textit{branch} $\in$ \textit{Branches\{\}}}
\If {\textsc{VerifyBranch(\textit{branch})} $\neq$ \textit{true}}
\State \textit{Branches} $\gets$ \textit{Branches \textbackslash branch}
\EndIf
\EndFor
\State \textit{Branches} $\gets$ \textit{SortByLength(Branches)}
\State \textit{Longest\{\}} $\gets$ \textit{SelectLongest(Branches)}
\If {$\vert \textit{Longest} \vert = 1$}
\State \Return \textit{Longest[0]}
\Else
\State \textit{Selected} $\gets$ \textit{Longest[0]}
\State \textit{counter} $\gets 1$ 
\While{\textit{counter} $< |$\textit{Longest}$|$}
\State \textit{current} $\gets$ \textit{Longest[counter]}
\State \textit{Divergent} $\gets$ \textit{GetFork(Selected,current)}
\If{\textit{LeaderAge(Selected, Divergent)} $<$ \textit{LeaderAge(current, Divergent)}}
\State \textit{Selected} $\gets$ \textit{current}
\EndIf
\If{\textit{LeaderAge(Selected, Divergent)} $=$ \textit{LeaderAge(current, Divergent)}}
\If{\textit{Binary(Selected,Divergent)} $<$ \textit{Binary(current, Divergent)}}
\State \textit{Selected} $\gets$ \textit{current}
\EndIf
\EndIf
\EndWhile
\State \Return \textit{Selected}
\EndIf
\EndProcedure
\end{algorithmic}
\end{algorithm}

\begin{algorithm}
\footnotesize
\caption{VerifyBranch}\label{MiningAlgorithmsVerifyBranch}
\begin{algorithmic}[1]
\Procedure{VerifyBranch}{\textit{currentBranch}}
\State $\textit{prevBlock} \gets \textit{Genesis(currentBranch)}$
\State $\textit{iterBlock} \gets \textit{Next(prevBlock, currentBranch)}$
\State $\textit{currentBlock} \gets \textit{Top(currentBranch)}$
\While{$\textit{iterBlock} \neq \textit{CurrentBlock}$}
\If {\textit{Hash(prevBlock)} $\neq$ \textit{prevHash(iterBlock)}}
\State \Return \textbf{false}
\EndIf
\State \textit{iterLeader} $\gets$ \textit{Leader(iterBlock)}
\If {\textit{iterLeader} $\notin$ \textsc{SelectCandidates(\textit{currentBranch[0, iterBlock]})}}
\State \Return \textit{false}
\EndIf
\State \textit{intent $\gets$ GetIntent(iterBlock)}
\If{\textit{GetTxHash(intent) $\neq$ Hash(GetTx(iterBlock))}}
\State \Return \textit{false}
\EndIf
\State \textit{counter} $\gets 0$
\ForAll{\textit{endorsement} $\in$ \textit{Endorsements(iterBlock)}}
\If {\textsc{VerifyEndorsement(\textit{endorsement,iterBlock})} $\neq$ \textit{true}}
\State \Return \textit{false}
\EndIf
\State \textit{counter} $\gets$ \textit{counter+1}
\EndFor
\If {\textit{counter} $<$ \textit{q}}
\State \Return \textit{false}
\EndIf
\If {\textit{VerifyVRF(iterBlock)} $\neq$ \textit{true}}
\State \Return \textit{false}
\EndIf
\ForAll{\textit{enrollment} $\in$ \textit{Enrollments(iterBlock)}}
\If {\textit{verify(enrollment)} $\neq$ \textit{true}}
\State \Return \textit{false}
\EndIf
\EndFor
\State \textit{prevBlock} $\gets$ \textit{iterBlock}
\State \textit{iterBlock} $\gets$ \textit{Next(iterBlock, currentBranch)}
\EndWhile
\State \Return \textit{true}
\EndProcedure
\end{algorithmic}
\end{algorithm}

\begin{algorithm}
\footnotesize
\caption{VerifyEndorsement}\label{MiningAlgorithmsVerifyEndorsement}
\begin{algorithmic}[1]
\Procedure{VerifyEndorsement}{\textit{endorsement, block, branch}}
\State \textit{endorser} $\gets$ \textit{GetEndorser(endorsement)}
\State \textit{intent} $\gets$ \textit{GetIntent(block)}
\State \textit{leader} $\gets$ \textit{GetLeader(block)}
\If{\textit{GetChainID(endorsement)} $\neq$ \textit{GetChainID(block)}}
\State \Return \textit{false}
\EndIf
\If{\textit{endorser} $\notin$ \textsc{SelectEndorsers(\textit{branch[0, block]})}}
\State \Return \textit{false}
\EndIf
\If{\textit{hash(intent)} $\neq$ \textit{GetIntentHash(endorsement)}}
\State \Return \textit{false}
\EndIf
\If{\textit{hash(leader)} $\neq$ \textit{GetLeaderHash(endorsement)}}
\State \Return \textit{false}
\EndIf
\State \textit{signature} $\gets$ \textit{GetSignature(endorsement)}
\State \textit{body} $\gets$ \textit{GetBody(endorsement)}
\State \Return \textit{VerifySignature(signature, body, endorser)}
\EndProcedure
\end{algorithmic}
\end{algorithm}


\section{Additional Parameter Values}
\label{app:extra-analysis}

In this appendix, we extend our analysis from Section~\ref{sec:analysis} to consider further example values for our system parameters.

\begin{figure*}[h]
  \centering
  \begin{subfigure}[b]{0.4\linewidth}
    \includegraphics[width=\linewidth]{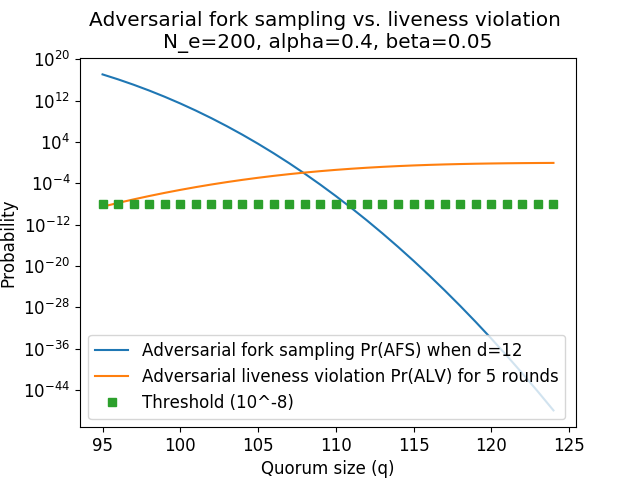}
    \caption{When $\alpha=0.4$ and $\beta=0.05$ using $N_e=200$ endorsers there is no quorum value $q$ that prevents forks at depth $d=12$ and provides good liveness.}
    \label{fig:a04}
  \end{subfigure}
  \hspace{10pt}
  \begin{subfigure}[b]{0.4\linewidth}
    \includegraphics[width=\linewidth]{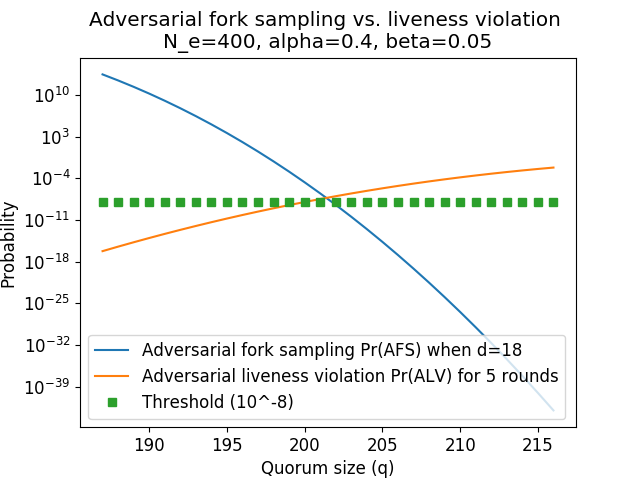}
    \caption{When $\alpha=0.4$ and $\beta=0.05$ using $N_e=400$ endorsers there is a quorum value $q=202$ that prevents forks at depth $d=18$ and ensures good liveness.}
    \label{fig:ne400a04}
  \end{subfigure}
  \par\bigskip
  \begin{subfigure}[b]{0.4\linewidth}
    \includegraphics[width=\linewidth]{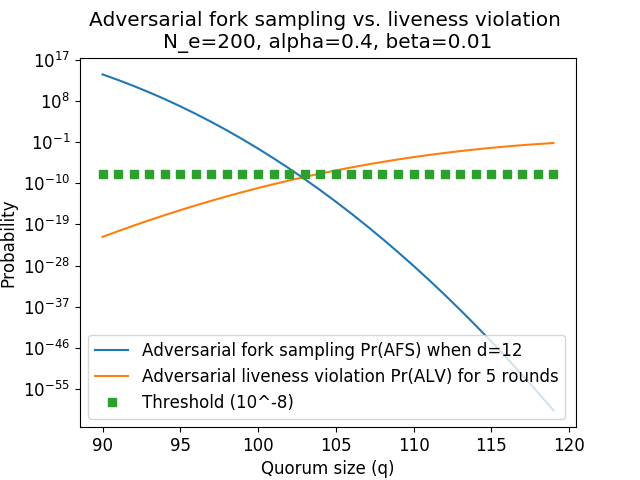}
    \caption{When $\alpha=0.4$ and $\beta=0.01$ using $N_e=200$ endorsers there is a quorum value $q=104$ that prevents forks at depth $d=12$ and provides good liveness.}
    \label{fig:a04b1}
  \end{subfigure}
  \hspace{10pt}
  \begin{subfigure}[b]{0.4\linewidth}
    \includegraphics[width=\linewidth]{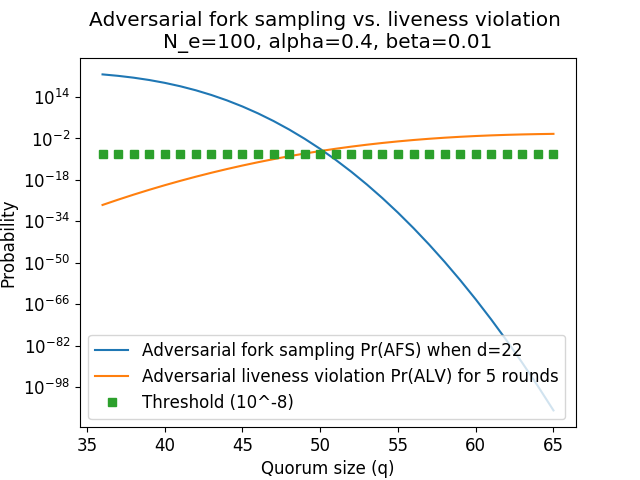}
    \caption{When $\alpha=0.4$ and $\beta=0.01$ using $N_e=100$ endorsers there is a quorum value $q=51$ that prevents forks at depth $d=22$ and provides good liveness.}
    \label{fig:a04b1ne100}
  \end{subfigure}
 \par\bigskip
  \begin{subfigure}[b]{0.4\linewidth}
    \includegraphics[width=\linewidth]{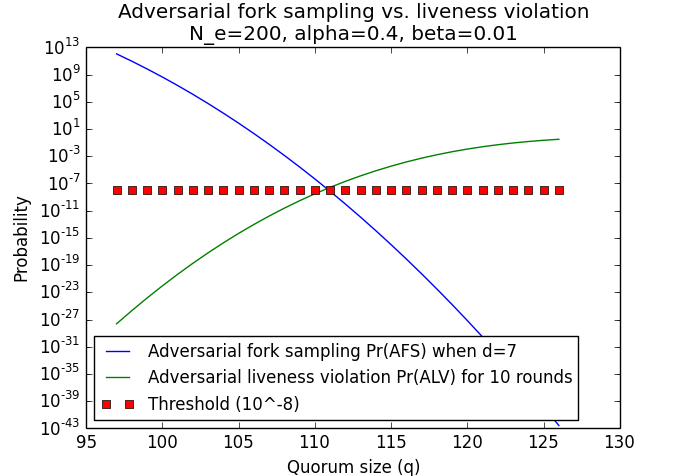}
    \caption{When $\alpha=0.4$ and $\beta=0.01$ using $N_e=200$ endorsers there is a quorum value $q=111$ that prevents forks at depth $d=7$, when adversarial liveness violation is increased to 10 rounds.}
    \label{fig:liveness1}
  \end{subfigure}
  \hspace{10pt}
  \begin{subfigure}[b]{0.4\linewidth}
    \includegraphics[width=\linewidth]{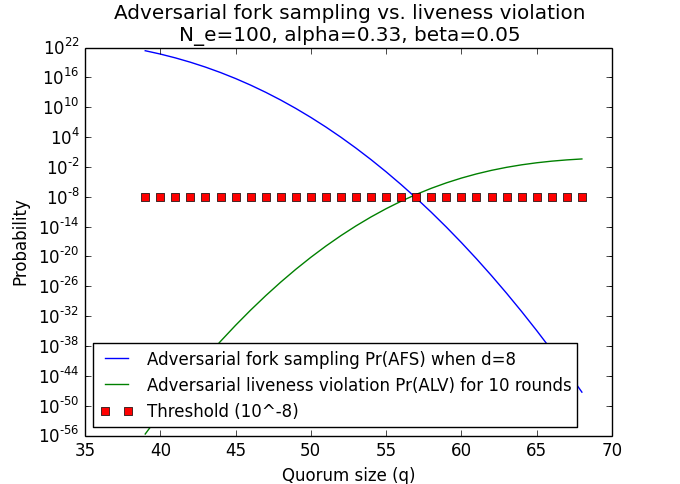}
    \caption{When $\alpha=0.33$ and $\beta=0.05$ using $N_e=100$ endorsers there is a quorum value $q=57$ that prevents forks at depth $d=8$, when adversarial liveness violation is increased to 10 rounds.}
    \label{fig:liveness2}
  \end{subfigure}
  \caption{Security versus liveness with additional example parameter values.}
  \label{fig:extra-plots}
  \vspace{-5pt}
\end{figure*}

We start by examining the effect of larger $\alpha$, i.e., cases where the adversary controls a larger fraction of all identities in the system. As can be see from Figure~\ref{fig:a04}, when $\alpha=0.4$ and the fraction of non-responsive identities remains as before ($\beta=0.05$), using our previous example value of $N_e=200$ endorsers, there is no quorum value $q$ that would prevent forks at the same depth $d=12$ without reducing liveness. To handle such cases we must either increase the endorser committee size or the maximum depth of forks. Figure~\ref{fig:ne400a04} shows that increasing the size of the endorser committee moderately to $N_e=400$ and simultaneously increasing the depth of the forks to $d=18$ allows us to find a quorum value $q=202$ that provides good security and liveness at the same time.

Tolerating such stronger adversaries ($\alpha=0.4$) becomes significantly easier in our solution when the connectivity between consensus nodes is better. As shown in Figure~\ref{fig:a04b1}, if we assume the fraction of non-responsive identities on each round to be smaller ($\beta=0.01$), it is possible to find a suitable quorum size ($q=104$) that provides prevents forks at depth $d=12$ using $N_e=200$ endorsers. Figure~\ref{fig:a04b1ne100} shows that in principle stronger adversaries ($\alpha=0.4$) can be handled without increasing the committee size ($N_e=100$) by only increasing the maximum depth of forks ($d=22$) which would mean a latency of almost two minutes.

Allowing longer adversarial liveness violation enables shallower forks. Next, we consider the case where we allow the adversary to prevent block creation for 10 rounds. Figure~\ref{fig:liveness1} shows that when $\alpha=0.4$ and $\beta=0.01$, using $N_e=200$ endorsers there is a quorum value $q=111$ that prevents forks at depth $d=7$ (in contrast to previous value $d=12$). Similarly, Figure~\ref{fig:liveness2} shows that when $\alpha=0.33$ and $\beta=0.05$, using $N_e=100$ endorsers there is a quorum value $q=57$ that prevents forks at depth $d=8$.

We conclude that our solution can handle various assumptions regarding the strength of the adversary and connectivity between the system participants, but our solution is best suited to scenarios where the adversary controls up to one third of all identities, but also stronger adversaries can be tolerated by using larger endorser committees or by reducing liveness guarantees.


\section{SGX Attestation Details}
\label{app:sgx}

\begin{figure}[t]
\centering
	\centering
	\includegraphics[width=0.9\linewidth]{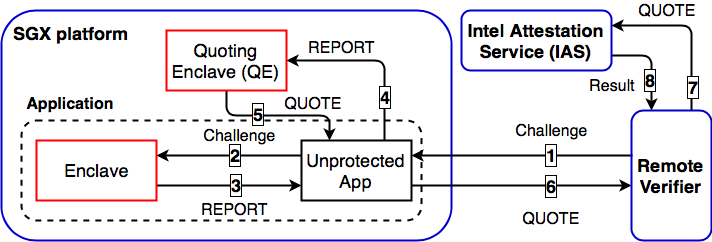}
	\caption{SGX remote attestation protocol that involves three parties: (i) the remote verifier, (ii) the attested SGX platform, and (iii) IAS that is an online service operated by Intel.}
 \label{fig:remote-attestation}
 \vspace{5pt}
\end{figure}

In this appendix, we provide further details on the SGX's attestation mechanism.
The enclave initialization actions performed by the OS are recorded securely by the CPU. This process creates a \emph{measurement} that captures the enclave's code configuration. Remote attestation is an protocol where an external verifier can verify that an enclave with the expected measurement was correctly initialized in a genuine SGX processor. The attestation protocol involves three parties: (i) the remote verifier, (ii) the attested SGX platform, and (iii) IAS that is an online service operated by Intel and it is illustrated in Figure~\ref{fig:remote-attestation}. 

The protocol proceeds as follows: (1) the remote verifier sends a random challenge to an unprotected application on the attested platform that (2) forwards it to the enclave that (3) returns a \textsf{REPORT} data structure encrypted for the Quoting Enclave containing the enclave's measurement. The \textsf{REPORT} data structure includes a \textsf{USERDATA} field, where the attested enclave can include application-specific attestation information, such as hash of its public key. (4) The application forwards \textsf{REPORT} to Quoting Enclave that (5) verifies it and returns a \textsf{QUOTE} structure signed by a processor-specific attestation key. (6) The application sends \textsf{QUOTE} to the remote verifier that (7) forwards it to the IAS online service that (8) verifies the \textsf{QUOTE} signature, checks that the attestation key has not been revoked, and in case of successful attestation returns the \textsf{QUOTE} structure signed by IAS. 

The attestation key is a part of a group signature scheme called Enhanced Privacy ID (EPID)~\cite{EPIDstandard} that supports two signature modes. The default mode is privacy-preserving. Another, linkable mode allows IAS to verify, if the currently attested CPU is the same as previously attested CPU. Usage of SGX attestation requires registration with Intel. Upon registration, each service provider receives a credential that they use to authenticate to IAS. If linkable mode of attestation is used, IAS reports the same \emph{pseudonym} every time the same service provider requests attestation of the same CPU~\cite{sgx_ias_api}.

\end{document}